\documentclass{ar2e}
\def\lesssim{\lower.7ex\hbox{${\buildrel < \over \sim}$}}
\def\gtrsim{\lower.7ex\hbox{${\buildrel > \over \sim}$}}
\def\mp{\lower.7ex\hbox{${\buildrel - \over +}$}}

\begin{document}

\input psfig.sty

\jname{Ann. Revs. Nucl. Part. Sci.}
\jyear{2002}
\jvol{52}
\ARinfo{1056-8700/97/0610-00}

\title{Flux of Atmospheric Neutrinos\footnote{
With permission from the Annual Review of Nuclear \& Particle Science. Final
version of this material is scheduled to appear in the Annual Review of
Nuclear \& Particle Science Vol. 52 , to be published in December 2002 by
Annual Reviews (http://annualreviews.org).
}}

\markboth{GAISSER and HONDA}{Flux of Atmospheric Neutrinos}

\author{T.K. Gaisser$^1$ and M. Honda$^2$
\affiliation{$^1$Bartol Research Institute, University of Delaware,
Newark, DE 19716 USA;\\
$^2$Institute for Cosmic Ray Research, University of Tokyo, 5-1-5 Kashiwanoha 
Kashiwa-shi, Chiba 277-8582, Japan}}

\begin{keywords}
atmospheric muons and neutrinos, cosmic rays, neutrino oscillations, 
neutrino astronomy
\end{keywords}

\begin{abstract}
Atmospheric neutrinos produced by cosmic-ray interactions in the atmosphere
are of interest for several reasons.  As a beam for studies of neutrino oscillations
they cover a range of parameter space hitherto unexplored by accelerator 
neutrino beams.  The atmospheric neutrinos
also constitute an important background and
calibration beam for neutrino astronomy and for the search for proton decay and other
rare processes.  Here we review the literature on calculations
of atmospheric neutrinos over the full range of energy, 
but with particular attention to the aspects
important for neutrino oscillations.  Our goal is to assess how well the properties
of atmospheric neutrinos are known at present.
 \end{abstract}

\maketitle

\section{Introduction}

When primary cosmic-ray protons and nuclei 
enter the atmosphere their interactions
produce secondary, ``atmospheric'' cosmic rays 
including all the hadrons and their
decay products.  The spectrum of these secondaries peaks in the GeV range, but
extends to high energy with approximately a power-law spectrum.
Because they are weakly interacting, neutrinos were the last
component of the secondary cosmic radiation to be studied, and
the time-scale for progress has been correspondingly slow.
The basic principles of detection and very rough estimates
of rates were known by 1960.  
Markov \cite{Markov} suggested the idea of detecting neutrino-induced
horizontal or upward-moving muons in a detector in a deep body of
water and estimated the rates that might be observed~\cite{MZ}.  
Greisen~\cite{Greisen} estimated roughly the rate of
neutrino interactions inside a large, deep Cherenkov detector.
Two groups~\cite{Achar,Reines} independently reported observations of 
atmospheric neutrinos in 1965 by 
detecting horizontal muons in detectors
so deep that the muons could not have been produced in the atmosphere.

With the advent of very large, deep underground detectors, whose construction
was originally motivated by the search for proton decay, there are
now extensive measurements of the flux of atmospheric neutrinos
in which the events occur within the fiducial volume of the detector.
What is emerging from recent studies is a discovery which is likely to have
an impact comparable to that of the discoveries made
with atmospheric cosmic rays half a century ago, 
including existence of the pion, the kaon 
and the muon.  The 50 kiloton SuperKamiokande experiment, with a 20 kiloton
fiducial volume, has now accumulated sufficient statistics to see
clearly that the flux of muon-neutrinos depends not only on the rate
at which they are produced in the atmosphere, but also on the pathlength
they travel to the detector~\cite{evidence}.  

The detector has a cylindrical shape, and its response
is up-down symmetric to within a few percent.  The 
atmospheric neutrino beam is also symmetric under reflection about
a horizontal plane through the detector--except for
geomagnetic effects that we discuss
in detail below.  
There is nevertheless a deficit of upward muon-neutrinos, with
pathlengths of order $10^4$~km, 
relative to downward neutrinos, which are produced within tens of kilometers
of the detector. 

The most likely cause of this pathlength-dependence is
neutrino oscillations, in which neutrinos produced as muon-type,
after passing a certain distance (which depends on energy),
manifest themselves with a different ``flavor", in this case
as $\nu_\tau$~\cite{SuperKtau}.  
The implication of this behaviour
is that the neutrinos possess a small but non-zero mass, thus
pointing to physics beyond
the standard model of particle physics.  This discovery
has stimulated intense activity to find the implications for
particle theory, and to relate this manifestation of neutrino
physics to the solar neutrino problem, which is now also understood to
be another manifestation of neutrino properties~\cite{SNO}.

The first evidence that atmospheric neutrinos were not behaving
as expected came from the observation by IMB that the fraction of
neutrino interactions with stopping muons was lower than expected~\cite{IMB1}.
Then Kamioka reported that the ratio of muon-like to electron-like
neutrino interactions was lower than expected~\cite{KAM}.  The Frejus experiment,
however, reported a ratio consistent with expectation ~\cite{Frejus}.  
These detectors
had fiducial volumes ranging from a fraction to a few kilotons.
As statistics accumulated from the water detectors 
(IMB and Kamioka)~\cite{IMB3a,IMB3b,KAM2,KAM3}
and results were reported from Soudan~\cite{Soudan}, 
evidence for the {\it atmospheric neutrino
anomaly} increased.  In the last paper from the Kamioka
experiment~\cite{KAMlast},
the first evidence for a pathlength dependence characteristic of
neutrino oscillations emerged.  

In the case of mixing between two flavors of neutrinos, one of which is
$\nu_\mu$, the probability that a neutrino produced as a $\nu_\mu$
travels a distance L and is detected as a $\nu_\mu$ is
\begin{equation}
\label{oscillation}
P_{\nu_\mu\rightarrow\nu_\mu}\;=\;
1\,-\,\sin^2(2\theta_m)\,\sin^2\left[1.27\,\delta m^2{L_{km}\over E_{GeV}}\right],
\end{equation}
where $\delta m^2 = m_1^2\,-\,m_2^2$ is the difference of the squared masses 
of the two related mass eigenstates and $\theta_m$ is the angle that
characterizes mixing between the two states.  Thus the characteristic
signature of neutrino oscillations is a deviation from the expected flux as
$L/E$ varies.  Currently, the best fit to the data of Super-K~\cite{SuperKtau}
has $\delta m^2\approx 3.2\times 10^{-3}$~eV$^2$ with full mixing,
$\theta_m\approx \pi/4$.

%\clearpage

\begin{figure}[!htb]%1
\centerline{\psfig{figure=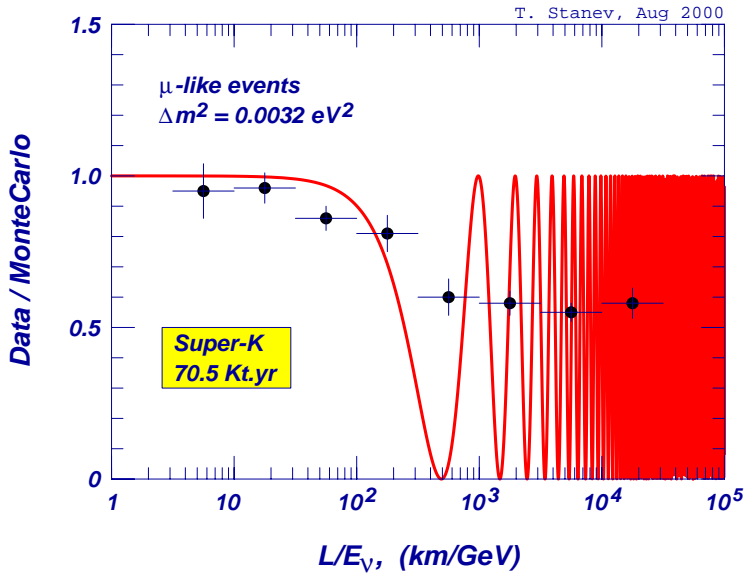,width=11.cm}}
\caption{\label{Fig1}Oscillation probability as a function of $L/E_\nu$.
}
\end{figure}

Fig.~\ref{Fig1} summarizes the current situation with
the measurements of Super-K.
One sees that the characteristic first oscillation dip cannot be resolved
at present with Super-Kamiokande.  This is a consequence of its
limited energy and angular resolution coupled with the fact that,
with the estimated value of $\delta m^2$,
the first minimum occurs for pathlengths
\begin{equation}
\label{critical}
L\;\approx\;400\, {\rm km}\times E_\nu(GeV).
\end{equation}
This distance from production in the atmosphere
to underground detector corresponds to neutrinos coming 
from near the horizon where the pathlength changes rapidly with zenith angle.
This consideration~\cite{Lipari1} limits the accuracy with which $\delta m^2$
can be determined with the atmospheric neutrino beam.
On the other hand~\cite{Lipari1} the determination of the value of the mixing
angle, since it is large, can be very precise with atmospheric neutrinos, 
limited only by statistics.

%%%%%%%%%%%%%%%%%%%%%%%%%%%%%%%%%%%%%%%%

\begin{figure}[!htb]%2
\centerline{\psfig{figure=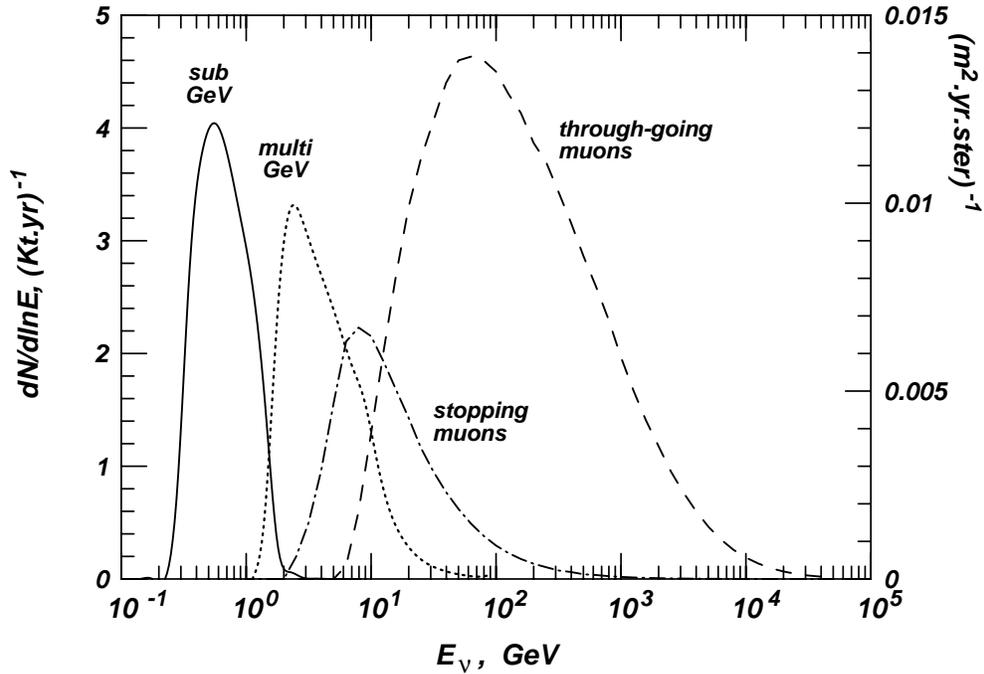,height=10cm}}
\caption{\label{response} 
Distributions of neutrino energies that give rise to
four classes of events.  Sub-GeV and multi-GeV refer to
the two classes of contained events at Superkamiokande.
Stopping and throughgoing muons refer respectively to
neutrino-induced muons produced outside the detector.
}
\end{figure}
%%%%%%%%%%%%%%%%%%%%%%%%%%%%%%%%%%%%%%%%

Contained neutrino events in Super-Kamiokande are classified as sub-GeV and 
multi-GeV.  To probe the atmospheric neutrino spectrum at higher energy it is 
necessary to make use the older technique of neutrino-induced muons,
where the effective volume is the projected area of the detector times
the muon range.  Since the muon range increases with energy, so does the
effective volume.  Fig.~\ref{response} shows the distribution of neutrino
energies for four classes of events.  
The $\nu_\mu\leftrightarrow\nu_\tau$ oscillation interpretation
that is emerging from the data is also consistent with deviations
from the expected angular distributions of neutrino-induced upward
muons measured in MACRO~\cite{MACRO} and 
Super-Kamiokande~\cite{SuperKupwardthrough}
and with the ratios of stopping to throughgoing upward 
muons~\cite{MACROstopping,SuperKupwardstopping}.   

The experimental evidence for oscillations of atmospheric neutrinos is
the subject to two recent reviews~\cite{KT,Jung}. 
Here we focus on calculations of the atmospheric neutrino flux.  
In doing so, we will emphasize those aspects 
that are important for making inferences about oscillations.
These include the flavor ratios and angular distributions
as well as pathlength distributions as a function of angle.

We will also discuss the absolute normalization of the flux and the
energy spectra up to very high energy.  While inferences about oscillations
emphasize measurements of ratios to minimize their dependence
on the absolute accuracy of atmospheric neutrino flux calculations,
the normalization is still important for checking the overall
consistency of the measurements and calculations.  
In addition, high energy
atmospheric neutrinos constitute the background for neutrino
astronomy at all energies.  Both Baikal~\cite{Baikal} and AMANDA~\cite{AMANDA}
have detected atmospheric neutrinos.  Neutrino telescopes with larger
effective volume are planned~\cite{Nestor,Antares,Nemo,IceCube}.  More important
is that the atmospheric neutrinos will be the primary calibration beam for
the neutrino telescopes.  

We organize the review as follows.  
After giving an overview of the calculations
in \S2, in the next three sections we
review the primary spectrum,
geomagnetic effects and hadronic interactions, which are the principal
physical quantities that influence the neutrino spectrum.
Then in \S6 we turn to a comparison of different calculations.
Until recently, nearly all calculations were one-dimensional in the sense
that all secondaries were assumed to follow the direction of the parent
cosmic-ray from which they descended.  In \S7 we review the novel features
of the recent three-dimensional calculations.  Atmospheric muons
are closely related to atmospheric neutrinos, and they therefore provide
an essential constraint.  We review the muon data in
relation to the neutrinos in \S8.  In \S9 we enumerate the uncertainties
in the various ingredients of the calculations and estimate the
resulting uncertainty in the atmospheric neutrino flux.  
 
\section{Overview of Neutrino Flux Calculations}

The atmospheric neutrino flux is a convolution of the primary spectrum at
the top of the atmosphere with the yield ($Y$)
of neutrinos per primary particle.  
To reach the atmosphere and interact, the primary cosmic rays first
have to pass through the geomagnetic field.
Thus the flux of neutrinos of type $i$ can be represented as
\begin{equation}
\phi_{\nu_i} \, = \,  \phi_p\,\otimes\,R_p\,\otimes\,Y_{p\rightarrow\nu_i}
 \, +\;\sum_A \left\{\phi_A\,\otimes\,R_A\,\otimes\,Y_{A\rightarrow\nu_i}
\right\},
 \label{nuflux}
\end{equation}
where $\phi_{p(A)}$ is the flux of primary protons (nuclei of mass A)
outside the influence of the geomagnetic field
and $R_{p(A)}$ represents the filtering effect of the geomagnetic field.
Free and bound nucleons are treated separately because propagation through
the geomagnetic field depends on magnetic rigidity (total momentum divided by
total charge) whereas particle production depends to a good approximation on
energy per nucleon.  A proton of
rigidity $R$~(GV) has total energy per nucleon 
$E(GeV)=\sqrt{R^2+m_p^2}$ whereas the corresponding relation 
for helium is $E(GeV/A)=\sqrt{R^2/4+m_p^2}$.

The neutrinos come primarily from the two-body decay modes
of pions and kaons and the subsequent muon decays.
The decay chain from pions is\begin{eqnarray}
\label{chain}
\pi^\pm\rightarrow & \mu^\pm & +\; \nu_\mu(\overline{\nu}_\mu)    \\
                   &\;\;\;\;\searrow &	\nonumber				        \\
 &  &e^\pm + \nu_e(\overline{\nu}_e)+\overline{\nu}_\mu(\nu_\mu),\nonumber
\end{eqnarray} with a similar chain for charged kaons.  When conditions are
such that all particles decay, we therefore expect\begin{eqnarray}
\label{features}
{\nu_\mu\, +\,\bar{\nu}_\mu\over\nu_e \,+\, \bar{\nu}_e} & \sim 2, & \\
\nu_\mu/\bar{\nu}_\mu\sim 1 \;&{\rm and}&\;
\nu_e/\overline{\nu}_e\sim\mu^+/\mu^-. \nonumber
\end{eqnarray}
Moreover, the kinematics of $\pi$ and $\mu$ decay is such that roughly
equal energy is carried on average by each neutrino in the chain.

\subsection{Early calculations}

The early calculations used the relation between muons and neutrinos
implied by Eq.~\ref{chain}.  The idea is to parameterize the
pion production spectrum in the atmosphere to fit an observed 
flux of muons.  In this way, the primary spectrum and the particle
production do not enter the calculation of the neutrino flux
explicitly.  Since the kinematical relation between neutrinos
and their parent pions is different for pions and kaons, some
assumption about their relative importance has to be made.

\begin{figure}[!thb]%3
\centerline{{\psfig{figure=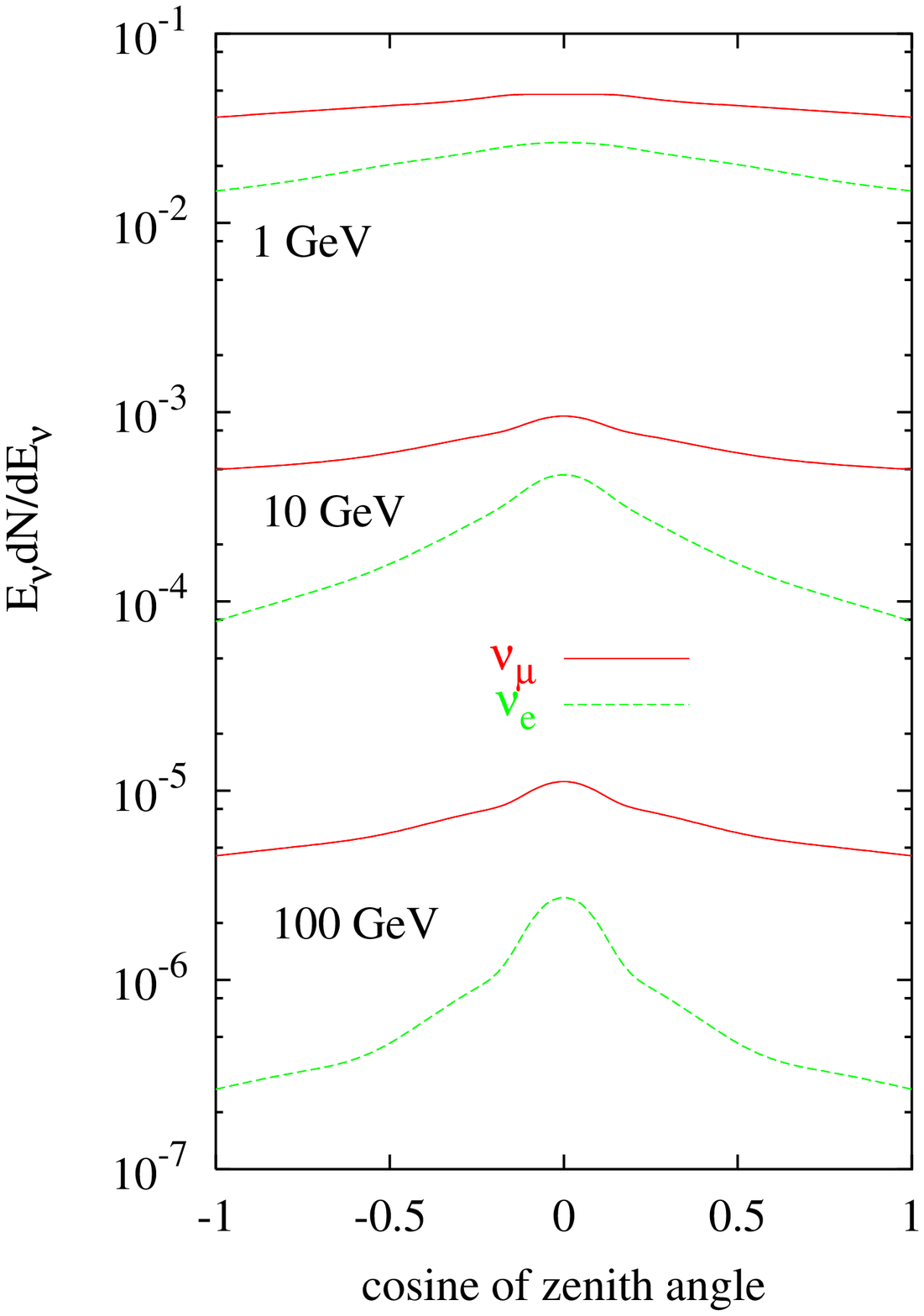,width=7cm,height=8cm} \hspace{9mm} 
{\psfig{figure=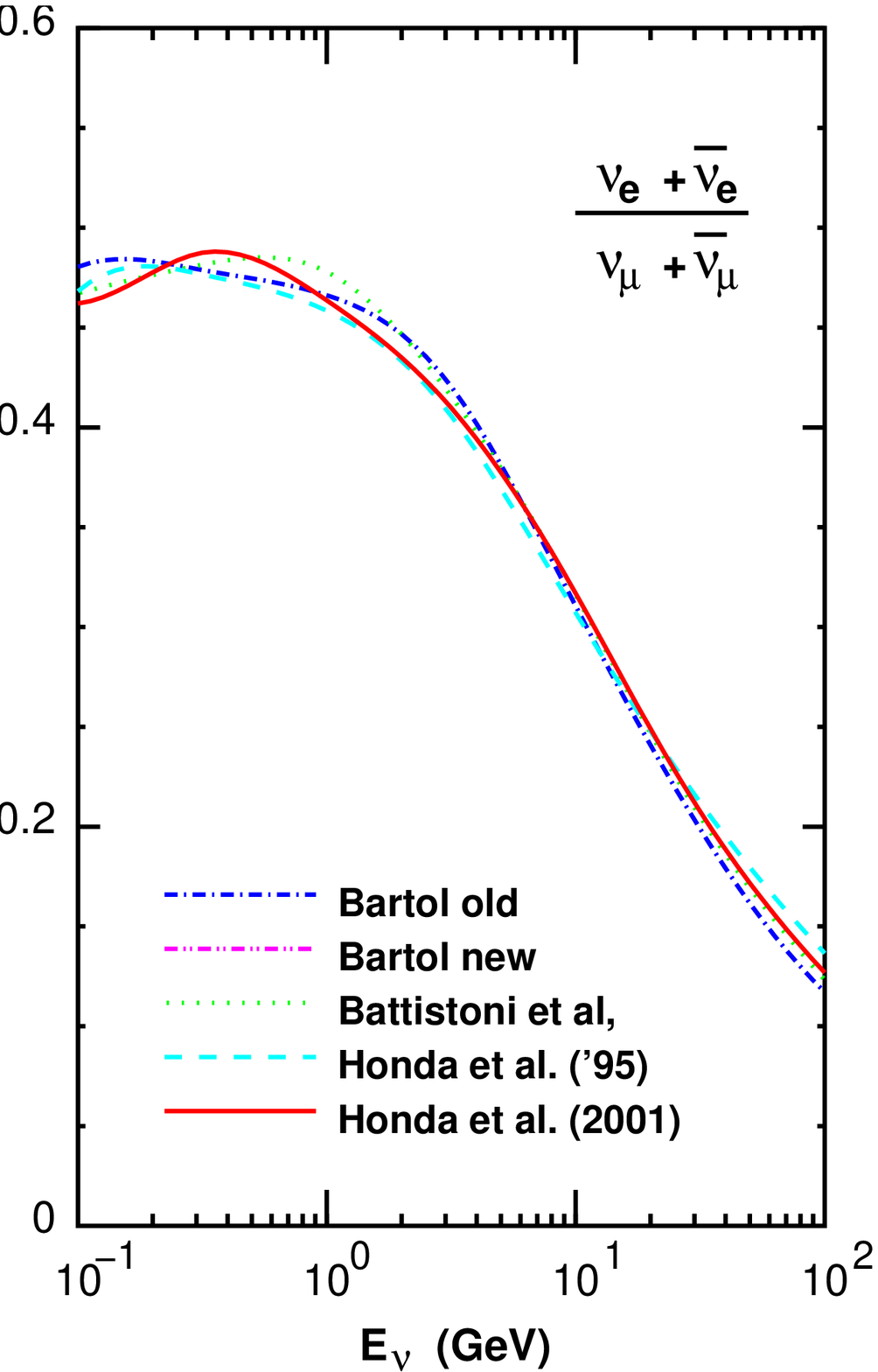,width=5.0cm}}}}
\caption{Left panel: Zenith-angle dependence of $\nu + \bar{\nu}$
calculated in the absence of a geomagnetic field.
Right panel: Ratio of $(\nu_e+\bar{\nu}_e)/(\nu_\mu+\bar{\nu}_\mu)$ as 
calculated in several recent papers including the geomagnetic
field for the location of Super-Kamiokande (see \S6).
}
\label{nuratio}
\end{figure}

Most of the essential features of the atmospheric neutrino spectrum were already
displayed and discussed in the 1961 paper of Zatsepin and Kuz'min~\cite{ZK}.
Muons with energy of several GeV and above reach the ground before decaying, so the
$\nu_e /\nu_\mu$ ratio decreases with increasing energy.  The resulting steepening of
the $\nu_e$ spectrum occurs at lower energy near the vertical direction and at 
higher energy near the  
horizon because of the longer decay path for the parent muons with large zenith angle.  
The component of the $\nu_\mu$ flux from muon decay behaves the same way.
For $E_\nu>100/\cos\theta$~GeV, the ratio of electron to muon neutrinos 
approaches a small value $\lesssim 5$\%, with the main source of $\nu_e$
the decay $K_L^0\rightarrow \pi\,e\,\nu_e$.  
To the extent that there is uncertainty
in kaon production, there will be a corresponding uncertainty in this ratio
at high energy.  These features of the neutrino flux are illustrated
in Fig.~\ref{nuratio}.  The left panel shows the neutrino fluxes from
Ref.~\cite{AGLS} calculated in the artificial case of no geomagnetic cutoff
and shown as a function of cosine of the zenith angle.  The right panel
shows the key ratio of electron neutrinos to muon neutrinos 
from several calculations~\cite{AGLS,Hamburg,Battistoni,Honda,Hondanew}
averaged over all directions for the specific location of Super-K.

In calculating neutrinos from decay of muons, it is important to include the
effect of muon polarization.  Although they did not include it in their
calculation, Zatsepin and Kuz'min pointed out that it would have the effect of
incrasing the $\nu_e/\nu_\mu$ ratio by about 5\%.  Muons are produced fully polarized
in the two-body decays of pions and kaons, with the direction of polarization
determined by the direction of the muon in the rest frame of its parent.
Because of the steep spectrum, the effect does not average to zero in the integral
over the parent energies.  Moreover, it has
the same sign for neutrinos and antineutrinos.  Volkova~\cite{Volkova} 
pointed out the importance of neutrino polarization in the context of the
atmospheric neutrino anomaly.

The angular and energy dependence of muons and neutrinos
from decay of pions and kaons is determined by the competition between
decay and interaction of the parent mesons.  The critical energy for pions in the
atmosphere is $\epsilon_\pi\approx 115$~GeV, 
and $\epsilon_K\approx 850$~GeV
for kaons.  For $E_\nu\ll\epsilon/\cos\theta$ the neutrino spectrum has approximately the
same power as the parent meson production spectrum and hence of the primary spectrum
of nucleons.  For $E_\nu\gg\epsilon\cos\theta$ the neutrino spectrum is 
one power steeper and is proportional to $\sec(\theta)$.
Since $\epsilon_K>\epsilon_\pi$, the spectrum of neutrinos from pion decay
steepens before that of neutrinos from decay of kaons, so that
kaons become increasingly important at high energy.  Since the change of slope
depends on zenith angle as well as energy,  it is possible in principle to use the observed
($E,\theta$) dependence of the muon flux
to estimate,  or at least set an upper limit, on the fraction of kaons.
Ref.~\cite{ZK1} used this method to show that atmospheric muons are not 
dominated by kaons.

The importance of kaons relative to
pions is further enhanced at high energies for neutrinos 
as compared to muons
because most of the energy in pion decay goes to the muon as compared to the
neutrino, whereas it is divided nearly equally in the $K\rightarrow\mu\nu$ decay.
For $E_\nu>100$~GeV kaons
become the dominant source of neutrinos even though 
the kaon/pion ratio at production is small.

Perkins~\cite{Perkins,Perkins1} has reviewed calculations 
that reconstruct the neutrino flux from the muon flux.  Refinements account
for variation of the geomagnetic field over the globe and start from muon data 
at high altitude where most of the pion and kaon decays occur.  
Recent calculations, however, use the method 
of Eq.~\ref{nuflux}, which is amenable to detailed monte carlo simulations, and we
will concentrate on those results in this review.  

\subsection{Analytic approximations}

Before proceeding to the details of the review, it will be helpful to 
discuss briefly the analytic approximations for the neutrino flux
as a guide to the physics of atmospheric neutrinos.  Not only do
these approximations allow a good qualitative understanding of
the features of the atmospheric neutrino flux, but they are valuable
for diagnostic purposes, for example, to determine what ranges of
primary energy and what regions of phase space in particle production
are important for various classes of events.  Often, for such diagnostic
purposes an approximate representation is adequate.  

\begin{figure}[!htb]%4
\centerline{\psfig{figure=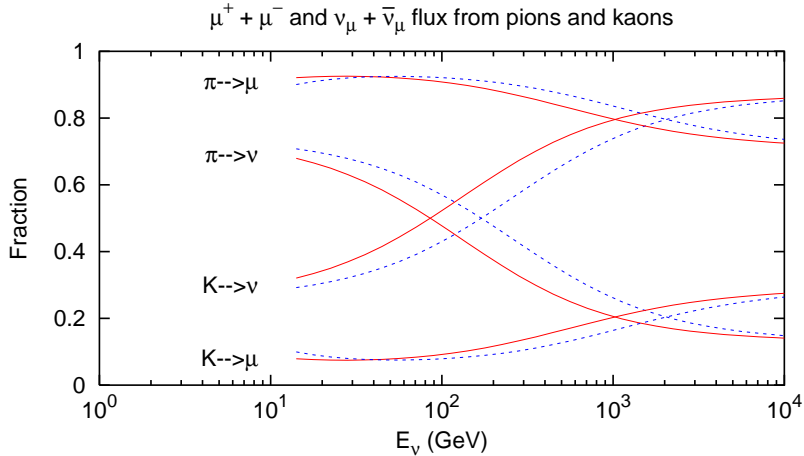,width=11cm}}
\caption{\label{PiK2}Fractional contribution of pions and
kaons to the flux of muons and neutrinos.  For pions the bottom curve of each
pair is for the vertical and the top curve 
for zenith angle of $60^\circ$.  The reverse is true for the kaon fractions.
}
\label{PiK}
\end{figure}

The analytic approximations apply at high energy ($E_\nu\gg 1$~GeV) and 
hold for a power-law primary spectrum and inclusive cross sections that
obey Feynman scaling in the fragmentation region.    With these assumptions,
the differential flux
of $\nu_\mu + \bar{\nu}_\mu$ from decay of pions and kaons is
\begin{equation}
{dN_\nu\over dE_\nu}\,=\,{\phi_N(E_\nu)\over (1 - Z_{NN})(\gamma+1)}\left\{
\left[{Z_{N\pi}(1-r_\pi)^\gamma\over1+B_{\pi\nu}\cos\theta E_\nu/\epsilon_\pi}\right] 
+ 0.635\left[{Z_{NK}(1-r_K)^\gamma\over1+B_{K\nu}\cos\theta E_\nu/\epsilon_K}\right]
\right\},
\label{analytic}
\end{equation}
where
\begin{equation}
\phi(E_0)\;=\;{dN\over dE_0} \;=\;A\times E_0^{-(\gamma+1)}
\label{power}
\end{equation} 
is the differential primary spectrum of nucleons of energy $E_0$.  
The neutrino flux in Eq.~\ref{analytic} is proportional to the
primary spectrum evaluated at the energy of the neutrino.
The constants $r_i=m_\mu^2/m_i^2$ for $i=(\pi,K)$, and the
constants $B_i$ depend hadron attenuation lengths
as well as decay kinematics~\cite{book}.
The expression for the flux of $\mu^++\mu^-$
has the same form but with different kinematical factors.
Similar (somewhat more complex) formulas also exist for neutrinos
from decay of muons~\cite{Paoloanalytic}.  
The approximations
become increasingly accurate as energy increases.
Eq.~\ref{analytic} displays the features of angular dependence
and the relative importance of kaons as described above and illustrated
in Fig.~\ref{PiK}.

\begin{table}[!htb]
\def~{\hphantom{0}}
\caption{Parameters for 
atmospheric $\nu_\mu +\bar{\nu}_\mu$ from $\pi$,K-decay}
%\centerline{
\begin{tabular}{@{}|lllll|@{}}
\toprule
Mass-square ratios&$r_\pi$&$r_K$&$B_\pi$&$B_K$ \\
\& B-factors:&0.573 & .046 &2.77 &1.18\\
\colrule
Characteristic $E_{decay}$:&$\epsilon_\pi$&$\epsilon_K$&$\epsilon_{charm}$& \\
&115 GeV & 850 GeV & $\sim 5\times10^7$ GeV & \\ \hline
Z-factors:&$Z_N$&$Z_\pi$&$Z_K$& \\
&0.30 &0.079 & 0.0118&  \\ 
\botrule
\end{tabular}
%}
\label{Tab1}
\end{table}

The physics of production of the parent pions and kaons
is contained in the spectrum-weighted moments,
\begin{equation}
Z_{N\rightarrow\pi(K)}\;=\;\int_0^1\,dx\,(x)^{\gamma - 1}\,
F_{N\rightarrow\pi(K)}(x),\;{\rm etc.},
\label{Zfactor}
\end{equation}
where
\begin{equation}
F_{N\pi}(E_\pi,E_N)
\,=\,{E_\pi\over\sigma_N}\,{d\sigma(E_\pi,E_N)\over dE_\pi}
\label{Zpion}
\end{equation}
is the normalized inclusive cross section for
$N\,+\,{\rm air}\rightarrow \pi^\pm\,+\,X$.
The integrand of the spectrum-weighted moment in Eq.~\ref{Zfactor}
displays the regions of longitudinal phase space in nucleon interactions
that are most important for production of neutrinos~\cite{response}.

Table 1 shows a sample set of parameters for the analytic formulas.  
The same formalism can be used to include neutrinos from decay of
charmed particles.  These eventually dominate the spectrum at sufficiently
high energy because of their short decay length (and correspondingly large
critical energy).

\subsection{Response functions}

%%%%%%%%%%%%%%%%%%%%%%%%%%%%%%%%%%%%%%%%
\begin{figure}[!htb]%5
\centerline{\psfig{figure=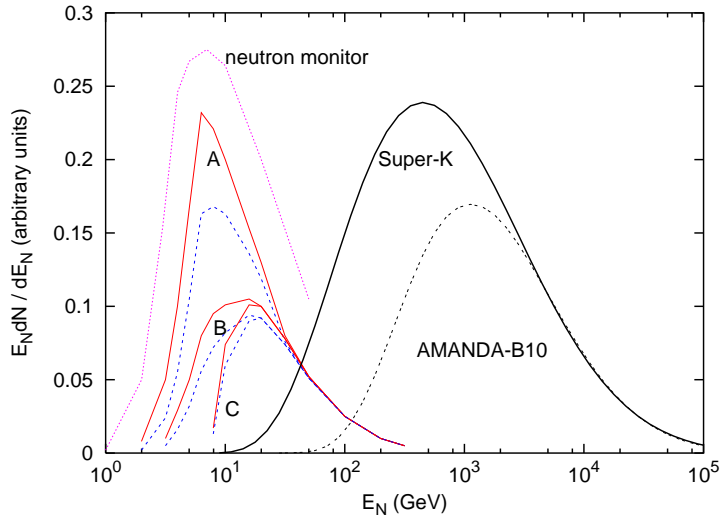,height=7cm}}
\caption{\label{response1} Response to primary energy
for several classes of 
interactions of $\nu_\mu+\bar{\nu}_\mu$: sub-GeV events
(A: no cutoff; B: events from the lower hemisphere;
C: events from upper hemisphere at Super-K); vertically upward,
throughgoing muons at Super-K and neutrino-induced muons in 
AMANDA-B10.  For the sub-GeV events,
each pair of curves shows the range of the signal between
minimum (solid) and maximum (dashed) level of solar activity.  The dotted
curve shows a typical response function for a neutron monitor,
to be discussed in \S3.3.  (The neutron monitor response is plotted
{\it vs} rigidity rather than energy.)
}
\end{figure}
%%%%%%%%%%%%%%%%%%%%%%%%%%%%%%%%%%%%%%%%

Response functions give the distributions of primary energy/nucleon
that produce neutrinos of a given energy (or distribution of energies).
Since the uncertainty in the primary spectrum depends on energy,
these functions can be used to evaluate the contribution of this
source of uncertainty to the calculated neutrino flux.

For the sub-GeV events the neutrino energies are low enough so that
geomagnetic cutoffs and solar modulation become
important.  These points are illustrated by the curves labelled
A, B and C in Fig.~\ref{response1} and will be
discussed further below.  Fig.~\ref{response1} also shows the response
for vertically upward throughgoing muons at Super-K
and for upward muons in AMANDA B10 as estimated with the approximations
of Ref.~\cite{response}.  The threshold for the neutrino-induced upward
muon at Super-K is $\approx 7$~GeV (vertical) and about an order
of magnitude higher for AMANDA B10~\cite{AMANDA}.

\section{Primary Spectrum}

Here we describe the primary spectrum inside the heliosphere
as it would be measured without the influence of the geomagnetic
field.  It is important to note that, except for geomagnetic effects,
the primary cosmic-ray flux is nearly isotropic.  Compton and Getting~\cite{Compton}
pointed out that the motion of the earth due to galactic rotation at $\sim 300$~km/s 
would lead to an asymmetry of order 1\% in the relativistic cosmic rays
if they there sources were external to the galaxy.  Any asymmetry due to the
motion of the earth relative to the cosmic-ray gas has come to be known as
the Compton-Getting effect.  It is now believed that most cosmic rays
originate inside the galaxy, so one might expect a smaller anisotropy
due to the peculiar motion of the solar system at $\sim 20$~km/s relative
to the galactic rotation.  Sidereal cosmic-ray 
anisotropies have been measured to be at the
level of $<0.1$\%~\cite{Groom}, although a recent analysis~\cite{Nagashima} 
indicates a more complex origin than the Compton-Getting effect.
In any case, such asymmetries 
are too small to affect the neutrino flux significantly.
We discuss the geomagnetic effects, which are important for
sub-GeV neutrinos, in the next section.

\subsection{Summary of data on protons and helium}

\begin{figure}[!htb]%6
\centerline{\psfig{figure=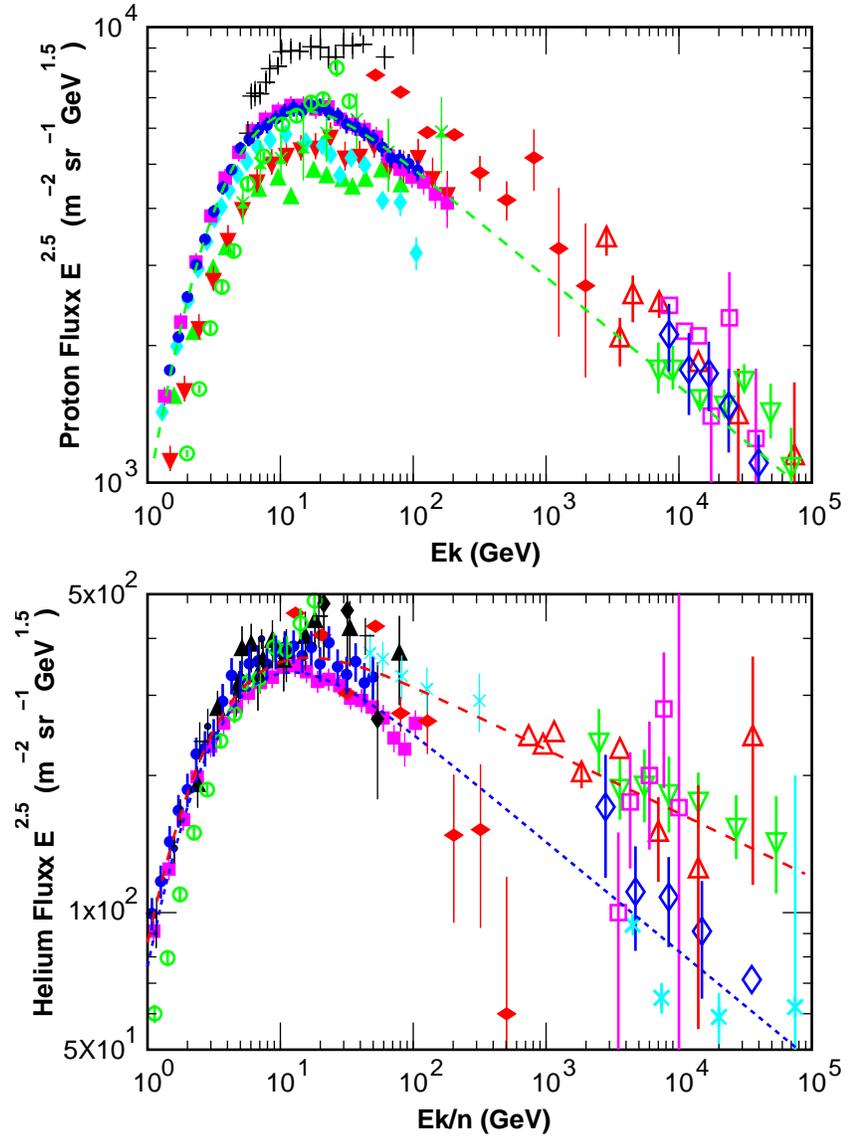,width=11.cm}}
\caption{\label{Fig-primary}
Observed flux of cosmic-ray protons and helium.
The dashed lines show the fits described in the text.  The data are:
Webber~\protect\cite{Webber}-crosses;
LEAP~\protect\cite{LEAP}-upward triangles;
MASS1~\protect\cite{MASS1}-open circles;
CAPRICE~\protect\cite{CAPRICE}-vertical diamonds;
IMAX~\protect\cite{IMAX}-downward triangles;
BESS98~\protect\cite{BESS}-circles; 
AMS~\protect\cite{AMS}-squares;
Ryan~\protect\cite{Ryan}-horizontal diamonds;
JACEE~\protect\cite{JACEE}-downward open triangles;
Ivanenko~\protect\cite{Ivanenko}-upward open triangles;
Kawamura~\protect\cite{Kawamura}-open squares;
Runjob~\protect\cite{Runjob}-open diamonds.
}
\end{figure}

In the last decade a series of measurements 
with superconducting magnetic spectrometers has greatly
improved our knowledge of primary cosmic-ray spectra up to
100 GeV/nucleon.  At higher energies the situation is not
as good.  Fig.~\ref{Fig-primary} is a summary of
the data for protons and helium.  The measurements below 100 GeV are made
with balloon-borne magnetic 
spectrometers~\cite{Webber,LEAP,MASS1,MASS,CAPRICE,IMAX,BESS}
and by the AMS Space Shuttle flight~\cite{AMS}.  Measurements at
higher energy have been made by balloon-borne calorimeters of various
kinds~\cite{Ryan,JACEE,Ivanenko,Runjob,Kawamura}.  
These do not capture all the energy of the primary, and as a
consequence, the energy determination is not as precise as with
a spectrometer.

Starting with the LEAP experiment~\cite{LEAP}, the measured fluxes
have been consistently lower that of Ref.~\cite{Webber}, which
had previously been the standard.  Particularly remarkable is
the close agreement between the BESS~\cite{BESS} and AMS~\cite{AMS}
measurements of the
proton flux.  They agree with each other to better than 5\% over
their whole energy range.  In contrast there is a systematic
difference of about 15\% between the BESS and AMS measurements
of helium.  Note, however, that only some 25~\% of all nucleons is carried by 
helium and heavier nuclei, so such a difference
corresponds to an uncertainty of $\sim$ 3~\% in the flux of nucleons.

\clearpage

Ref.~\cite{spectrum} discusses the measurements of the
primary spectrum in the context of atmospheric neutrinos.
Fits of the form
\begin{equation}
\phi(E_k)\;=\;K\times
\left(E_k\,+b\,\exp\left[-c\sqrt{E_k}\right]\right)^{-\alpha}
\label{Todor}
\end{equation}
are given, 
with parameters as tabulated in Table~\ref{Tab2}.  The fits 
for protons and helium are 
determined largely by the AMS~\cite{AMS} and BESS~\cite{BESS}
data, with their small statistical uncertainties.  They are
shown as dashed lines in Fig.~\ref{Fig-primary}.
Parameters for the three groups of heavy nuclei
in Table~\ref{Tab2} have been updated by Stanev~\cite{Stanev-private} 
since publication of Ref.~\cite{spectrum}.

\begin{table}[!htb]
\def~{\hphantom{0}}
\caption{Parameters for all five components 
in the fit of Eq. \protect\ref{Todor}.}
\vspace*{2truemm}
\begin{tabular}{@{}|l| l l l l|@{}}% \hline
\toprule
 parameter/component  & $\alpha$ &  K & b & c  \\ % \hline
\colrule 
 Hydrogen (A=1) & 2.74$\pm$0.01 & 14900$\pm$600 & 2.15 & 0.21 \\
 He (A=4, high) & 2.64$\pm$0.01 & 600$\pm$30  & 1.25 & 0.14 \\ 
He (A=4, low) & 2.74$\pm$0.03 & 750$\pm$100  & 1.50 & 0.30 \\ 
 CNO (A=14)  & 2.60$\pm$0.07 & 33.2$\pm$5 & 0.97 & 0.01 \\
 Mg-Si (A=25)  & 2.79$\pm$0.08 & 34.2$\pm$6  & 2.14 & 0.01 \\
 Iron (A=56)  & 2.68$\pm$0.01  & 4.45$\pm$0.50 & 3.07 & 0.41 \\ %\hline 
\botrule
  \end{tabular}
% \end{center}
\label{Tab2} 
\end{table}

\subsection{Heavier nuclei and the all-nucleon spectrum}

%%%%%%%%%%%%%%%%%%%%%%%%%%%%%
\begin{figure}[!htb]%7
\centerline{\psfig{figure=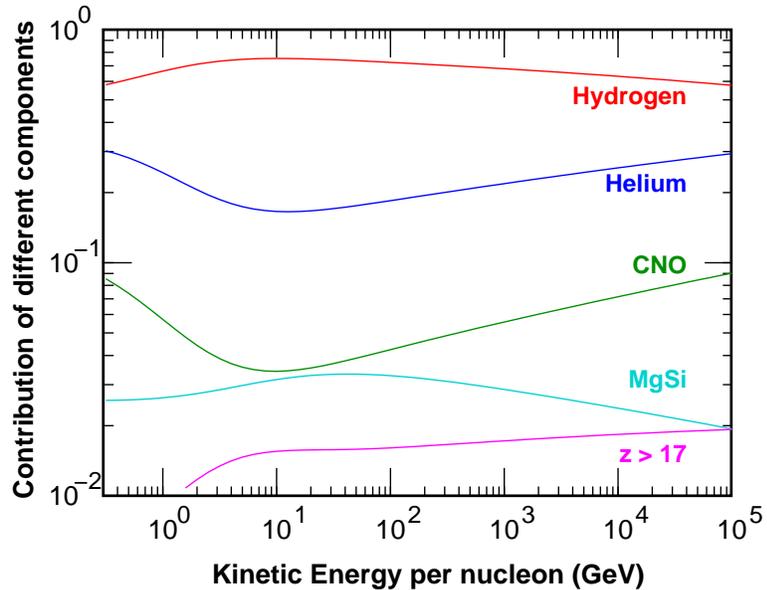,width=10.cm}}
\caption{\label{fractional}
Fractional contributions of various nuclei to
the all-nucleon spectrum~\protect\cite{spectrum}.}
\end{figure}
%%%%%%%%%%%%%%%%%%%%%%%%

Production of pions and kaons in the atmosphere, and 
hence, the neutrino flux, depends essentially on the spectrum of nucleons
as a function of energy-per-nucleon.  Protons and neutrons must
be treated separately to obtain the 
charge ratio and the correct  $\bar{\nu}/\nu$
ratios.  Bound and free protons must be passed separately through
the geomagnetic filter since the relation between energy per nucleon
and rigidity is different for free protons and for nuclei.

A fit for the all-nucleon spectrum outside the magnetosphere
corresponding to Eq.~\ref{Todor} can be constructed
directly from Table~\ref{Tab2}.  Fig.~\ref{fractional}
shows the fractional contributions of protons, helium
and three groups of heavy nuclei.  In the energy range
important for the contained neutrino interactions, protons
contribute 75\%, helium 15\% and all heavier nuclei
about 10\%.  

\subsection{Solar modulation}

%%%%%%%%%%%%%%%%%%%%%%%%%%%%%
\begin{figure}[!htb]%8
\centerline{\psfig{figure=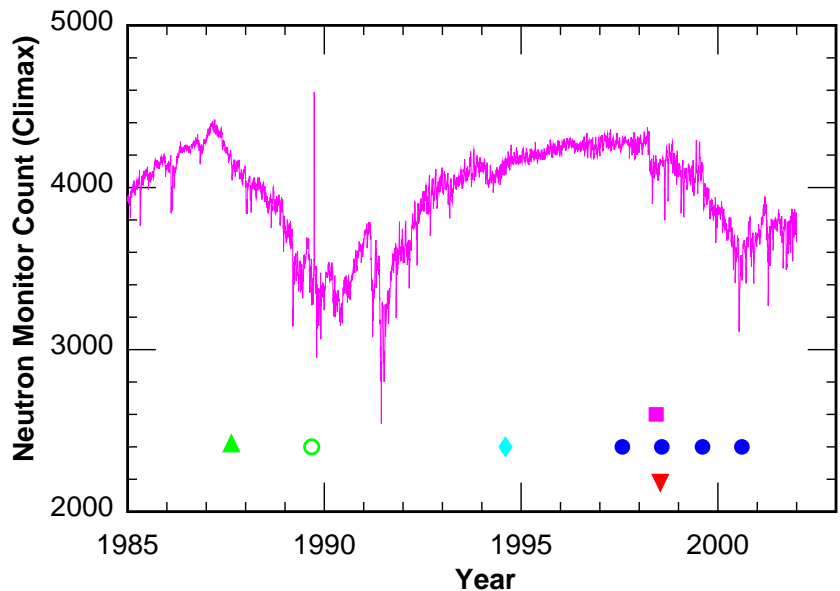,width=11.cm}}
\caption{\label{modulation}
Record of count rates for the Climax neutron monitor~\cite{Climax}.  
Times of the 
various flights are marked using the same symbols as in 
Fig.~\protect\ref{Fig-primary}.
}
\end{figure}
%%%%%%%%%%%%%%%%%%%%%%%%

The sun emits a magnetized plasma with a velocity of $\approx400$~km/s
in the solar equatorial region and about twice as high over the
solar poles~\cite{Ulysses}.  To reach the earth and interact
in the atmosphere, galactic cosmic rays have to diffuse into
the inner heliosphere against the outward flow of the turbulent
solar wind, a process know as solar modulation.  Particles of 
very low energy outside the heliosphere are nearly completely
excluded, while higher energy particles lose energy as they diffuse in.
During periods
of high solar activity (``solar maximum'') the turbulence in the
solar wind is higher than during solar minimum and the low-energy
portion of the spectrum is more suppressed.

A classic tool for study of solar modulation is the neutron 
monitor~\cite{Simpson}.  Response functions for neutron
monitors~\cite{Clem} are similar to the response for sub-GeV neutrinos,
as illustrated by the dotted curve in Fig.~\ref{response1}.
As a consequence, neutron monitor records~\cite{Bartol,Climax} 
provide an ideal
tool for interpolating between the spectra measured at particular
instants of the solar cycle.  Fig.~\ref{modulation} is a plot of
the Climax neutron monitor~\cite{Climax} 
with the dates of the flights on which the
low-energy data in Fig.~\ref{Fig-primary} were obtained.
Except for MASS1~\cite{MASS1}, most of the flights
were close to solar minimum conditions, and the shapes of their low-energy
spectra are similar.  The MASS1 data, which were obtained in 1989 during
solar maximum conditions, show extra suppression at low energy.  

%%%%%%%%%%%%%%%%%%%%%%%%%%%%%
\begin{figure}[!htb]%9
\centerline{\psfig{figure=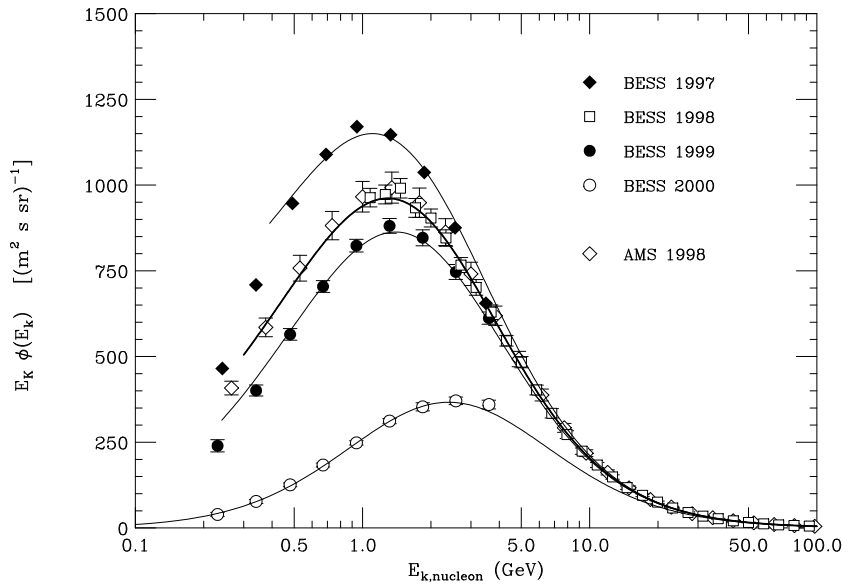,width=12.cm}}
\caption{\label{BESS4}
Proton spectra measured by BESS in 1997, 1998, 1999 and 
2000~\protect\cite{Asaoka}.  Curves are explained in the text.
}
\end{figure}
%%%%%%%%%%%%%%%%%%%%%%%%

It is possible to use the neutron monitor count rate ($N_m$) as a parameter
to relate the spectrum at one phase of the solar cycle to that at
another.  The parameterization of Ref.~\cite{Nagashimaetal} was used in
Ref.~\cite{Honda} to provide a multiplicative
factor by which a time-independent reference spectrum is multiplied.
The factor is a function both of particle momentum and $N_m$.
Recently a new set of
data taken with the BESS detector at three different levels of solar activity
(including one in 2000 during solar maximum) has been published~\cite{Asaoka}.
 We show these data, together with the BESS98 data from Fig.~\ref{Fig-primary},
 in Fig.~\ref{BESS4}.
The curves in Fig.~\ref{BESS4} are obtained by 
Lipari~\cite{Paolo-private}
starting with Eq.~\ref{Todor} for
the 1998 BESS data and using the 
force-field approximation~\cite{forcefield}.  In this simple
approximation the phase space density of particles is shifted
down in energy by an amount $\Delta\Phi$ that depends only on $N_m$ and not
on momentum.  When the BESS-2000 data for $E>4$~GeV becomes available it will
be possible to refine the way the spectrum modulates as a function of
energy.  At the same time, the data
should provide the basis for an improved understanding of solar modulation.
In the meantime, we can estimate roughly (see Fig.~\ref{response1})
that solar modulation should be a 10\% effect at Kamioka
and a 20\% effect at the high latitude sites, Soudan and Sudbury.

\section{Geomagnetic effects}
\label{geomag}

The geomagnetic field affects cosmic rays both outside and
inside the atmosphere. 
Outside it acts as a
filter which allows in particles of sufficiently high energy 
and excludes those of lower energy.  Inside it bends charged secondaries
causing some important effects which we discuss later.
Whether a particle is allowed or forbidden is determined by its
position, direction, and radius of curvature.  Only those particles
that interact in the atmosphere before curving back into
space can contribute to the flux of atmospheric neutrinos.  
Because the effect depends on the gyroradius of a particle,
the relevant kinematic variable is total momentum divided by total
charge, i.e. rigidity.

In case of a dipole magnetic field centered on the earth, 
the cutoff rigidity can be
expressed in an analytic form by St\"ormer's formula~\cite{Stormer,Alpher}:
\begin {equation}
R_S^\pm (r, \lambda_M, \theta, \varphi) =  \left ( {M \over  r^2 }
\right )  ~ \left \{  { \cos^4 \lambda_M \over
[1 + (1 \mp \cos^3 \lambda_M \sin \theta \sin \varphi)^{1/2}]^2 } \right \},
\label{eq:Stormer}
\end{equation}
where $r$ is the distance from the center of the earth and 
$\lambda_M$ is the geomagnetic latitude, 
$\theta$ and $\varphi$ define the arrival direction of the cosmic ray,
and $M=8.1\times 10^{25}$~G\,cm$^3$ 
is the magnetic dipole moment of the earth.
The azimuthal angle of the direction vector of the
particle is $\varphi$, measured counterclockwise
from magnetic north.  Thus a particle arriving from the west
has $\sin\varphi < 0$ giving a lower cutoff for a positive
particle (upper sign in Eq.~\ref{eq:Stormer}).  A positive particle
arriving from the east has a higher cutoff.  The scale
is set by the maximum cutoff $eM/r_\oplus^2 =60$~GeV, which 
is the energy at which a proton approaching from the
east at the equator will orbit the earth.

\subsection {Cutoffs for realistic magnetic field}

The actual magnetic field is more complicated than a symmetric dipole,
and such a simple expression as Eq.~\ref{eq:Stormer}
is not available.
The geomagnetic field is generally expressed as a multipole 
expansion, and its global structure is well known~\cite{geomagfield}.

The cutoff rigidity for a realistic geomagnetic field can be calculated
by the back tracing technique.  An anti-proton,
which has the same mass as a proton but the opposite charge, 
is used as the test particle.
We note that the change $e \leftrightarrow -e$ is equivalent to
the change of $t \leftrightarrow -t$ in the equation of motion
of a charged particle in a magnetic field.  %;
%\begin{equation}
%{\partial {\bf p} \over \partial t} = e {\bf v} \times {\bf B}
%\label{eq:eq-motion},
%\end{equation}
%where {\bf p} is the momentum, 
%{\bf v} is the velocity,
%and {\bf B} is the magnetic field.
We launch anti-protons from the top of the atmosphere of the Earth, 
varying the position and direction to obtain
the coverage needed for a given detector location.
Standard practice is to define the escape sphere
at 10 earth radii. 
The rigidity cutoff is calculated as the
minimum momentum with which the test particle escapes from the
geomagnetic field.  For some directions at some locations the
situation is still more complicated.  There can in some cases
also be forbidden trajectories in a narrow region 
just above the minimum cutoff~\cite{LScut}.

\begin{figure}[!thb]%10
\centerline{\psfig{figure=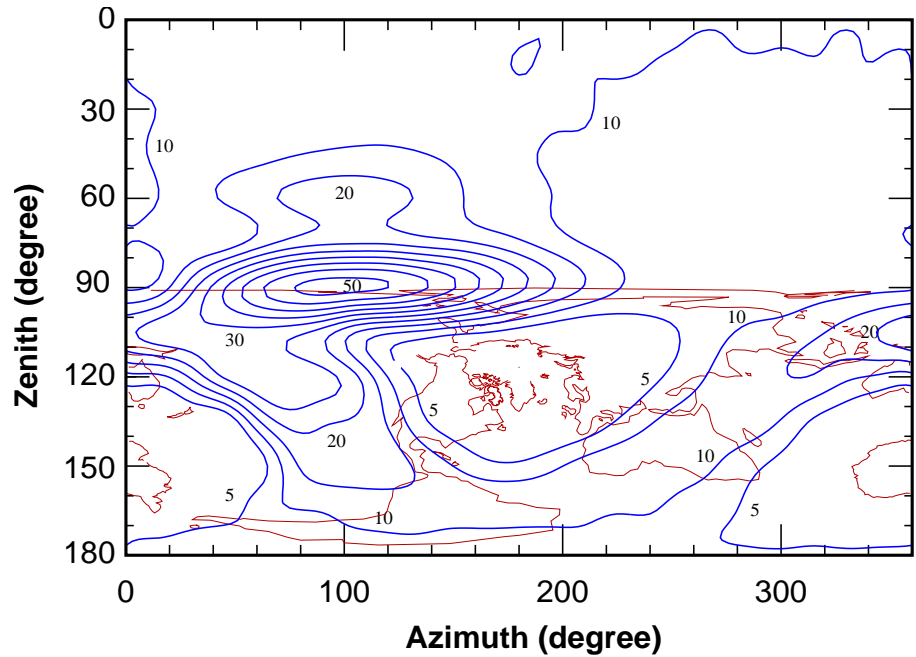,width=12.cm}}
\caption{Contour map of the rigidity cutoff as seen from the Kamioka site,
Rigidity cutoffs are shown as a function of arrival direction of
the neutrino. (Zenith angle $>90^\circ$ is for upward-moving particles.)
An outline map of the continents is
superimposed on the lower hemisphere.}
\label{Fig-geom-contour}
\end{figure}

The back tracing technique only tells us if there is a path along 
which a cosmic ray of a particular momentum and charge
can reach the top of the atmosphere.
To know the spectrum at the 
top of the atmosphere, however, one also needs Liouville's theorem,
which ensures the conservation of particles in phase space.
Then, assuming that the distribution of cosmic rays outside
the heliosphere is isotropic and assuming no change
in magnitude of momentum along the path, the flux of particles
at the atmosphere has the same value as in interstellar space
for allowed values of momentum.

\begin{figure}[!thb]%11
\centerline{{\psfig{figure=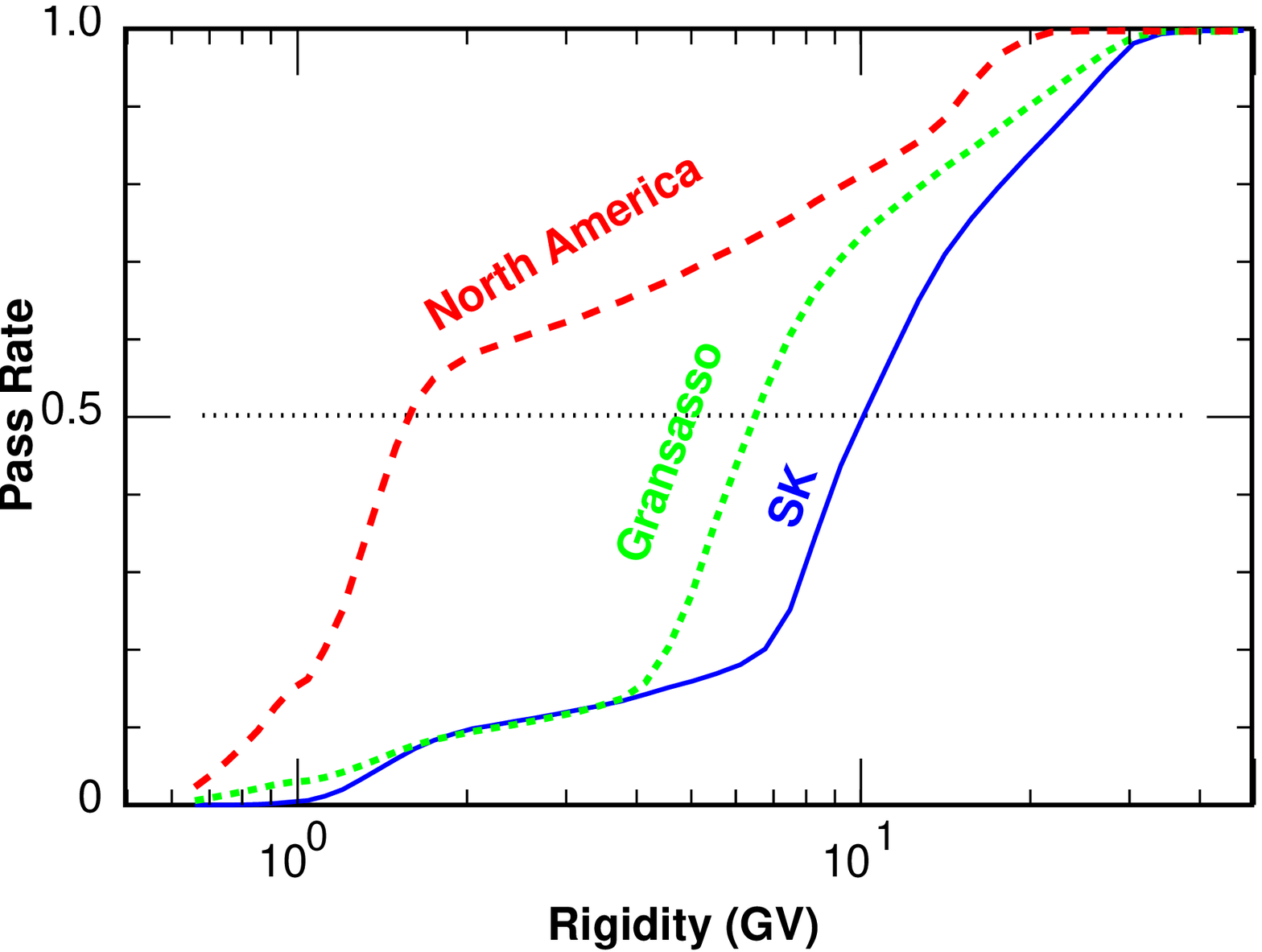,width=6.cm} \hspace{5mm}
{\psfig{figure=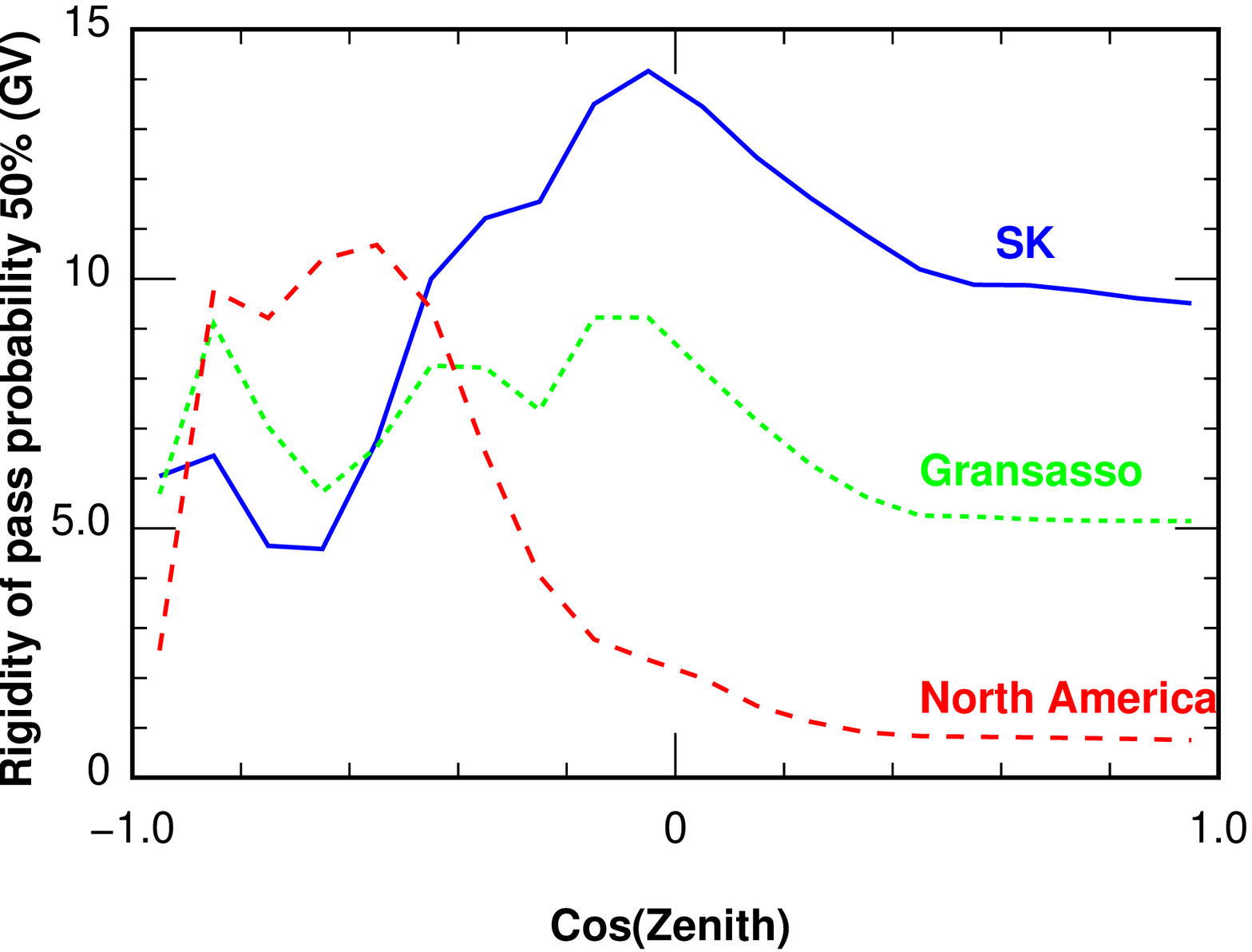,width=6.cm}}}}
\caption{
Cutoffs at three locations: north-central North America (Soudan, Sudbury);
Gran Sasso; Kamioka.  The left panel shows the passing rate as a function of
rigidity integrated over all directions.  The right panel shows the rigidity
above which half the particles in the azimuthal band at each zenith angle
reach the atmosphere.  (Positive $\cos\theta$ refers to downward-moving particles.)
}
\label{passrate}
\end{figure}

For a one-dimensional calculation, one can make
a contour map of the cutoffs for each detector 
as a function of direction along the line of sight
from the detector.  Each entry is the cutoff in the atmosphere where
the neutrino trajectory originates.
Such a contour map for the cutoffs
at Kamioka is shown in Fig.~\ref{Fig-geom-contour}~\cite{Honda}.
In the case of a three dimensional calculation, the treatment of geomagnetic 
cutoffs is very considerably more complex.  Since the direction
of the neutrino can be at an angle relative to the primary cosmic
ray that produced it, one needs the cutoffs for all directions
at every point around the globe.

The qualitative effects of the geomagnetic cutoffs at different
locations on the globe can be displayed by comparing averaged cutoffs.
The left panel of Fig.~\ref{passrate} plots the fraction 
of $4\pi$~sr allowed as a function of rigidity.
The second panel shows the cutoffs averaged over
azimuth plotted as a function of zenith angle. (The average
cutoff for a given zenith is defined as the rigidity at
which 50\% of the directions in the azimuth band are allowed.)  
The three locations
are North America (Sudbury and Soudan), Gran Sasso and Kamioka
in increasing order of average cutoff.  While the average cutoffs
for the lower hemisphere ($\cos\theta < -0.5$) are similar for all locations
(because they involve averaging over 
large portions of the surface of the earth)
the local cutoffs are quite different.  In particular,
the up-down ratio at Super-K is opposite to that at the high-latitude
sites in North America.  Moreover, the geomagnetic up-down ratio
at Super-K is such as to suppress downward events, which is opposite to
the observed pathlength dependence.

\subsection {East-West Effect}

Kamioka is at a rather low geomagnetic latitude and has therefore
a rather high local cutoff, $\approx11$~GV.  
In addition, there is a strong azimuthal
asymmetry, with the cutoff higher for particles arriving from
the east than from the west.  The east-west asymmetry follows from
the fact that the primary cosmic rays are positively charged.  
Historically, it was the east-west asymmetry that led to the conclusion
that the primary cosmic rays are positively charged~\cite{Johnson,Alvarez}.
A remarkable feature of the Super-K is that it has
been possible to measure~\cite{SuperKEW} the classic east-west asymmetry of the
primary cosmic rays 
as reflected in neutrinos~\cite{LSG,Honda}.  
Quantitative understanding
of this feature of the data confirms that deviations of
the atmospheric neutrino beam from isotropy can be understood
entirely as arising from effects of the geomagnetic
field.  

\begin{figure}[!thb]%12
\centerline{\psfig{figure=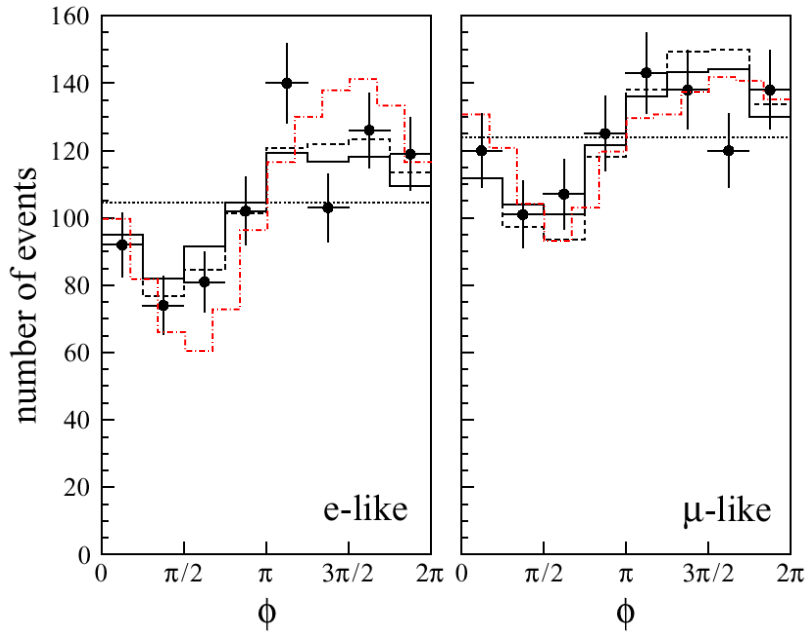,width=11.cm}}
\caption{East West effect observed by neutrino.
Data points are from SK experiment, and solid and dashed lines are 
prediction by using the one-dimensional calculation of Honda et al. 
and Bartol group.
Dash-dot are predictions by Lipari taking into account the 
muon bending normalized by the average value. }
\label{Fig-sk-ew}
\end{figure}

In the Fig.~\ref{Fig-sk-ew}, we show the experimental data with 
the predictions. It is seen that the experimental data agree with the 
predictions well within the experimental errors.
However, Lipari pointed out that the azimuthal variation of 
$\nu_e$ events is somewhat larger than the predictions
of the one-dimensional calculations, while the azimuthal variation
of the $\nu_\mu$ events is smaller~\cite{Lipari-ew}.
He  argues that these details are a consequence of bending of
muons by the geomagnetic field inside
 the atmosphere, which is neglected in the one-dimensional calculations.

Roughly speaking, the $\mu^+$ bending enhances the E-W effect for the 
primary cosmic rays, since the $\mu^+$ has the same charge as primary
cosmic rays. 
On the other hand the $\mu^-$ bending works to cancel the E-W effects.
Therefore the azimuthal variation of $\nu_e$ and $\bar\nu_\mu$ is 
enhanced by the muon bending above the prediction of the one-dimensional
calculations. On the other hand, the azimuthal variation of $\bar\nu_e$ 
and $\nu_\mu$ are reduced.
Considering the fact that the interaction cross sections for $\nu$'s 
are $\sim$ 3 times larger than $\bar\nu$'s at $\sim$ 1~GeV, 
the muon bending enhances the variation of electron events in the 
detector and reduces the variation of muon events.

In Fig.~\ref{Fig-sk-ew} the data are plotted as a function of 
azimuth for a fixed band of zenith angles.  Thus any pathlength-dependence
affects all events in the band in the same way.
The fact that the observed azimuthal variation can be understood entirely 
as a consequence of geomagnetic effects
therefore confirms that the observed pathlength
dependence (i.e. dependence on zenith angle) is not due to a poorly
understood geomagnetic effect, but is a consequence of pathlength-
dependence of the neutrinos.

\subsection{Second (subcutoff) spectrum}

The AMS detector on the Space Shuttle found and measured~\cite{AMS2}
at 380 km altitude a population of partially trapped particles
below the local geomagnetic cutoff.  This population is related
to the cosmic-ray albedo previously known from stratospheric balloon flights
 above 30 kilometers (see, for example,~\cite{albedo,albedo2}).  
It originates when higher energy
cosmic rays above the local geomagnetic cutoff interact in
the atmosphere and produce secondaries that curve away from the earth.
Some of the secondaries may be below the local geomagnetic cutoff, 
in which case they eventually re-enter the atmosphere as
re-entrant albedo~\cite{reentrant}.

%%%%%%%%%%%%%%%%%%%%%%%%%%%%%%%%%
\begin{figure}[!thb]%13
\centerline{\psfig{figure=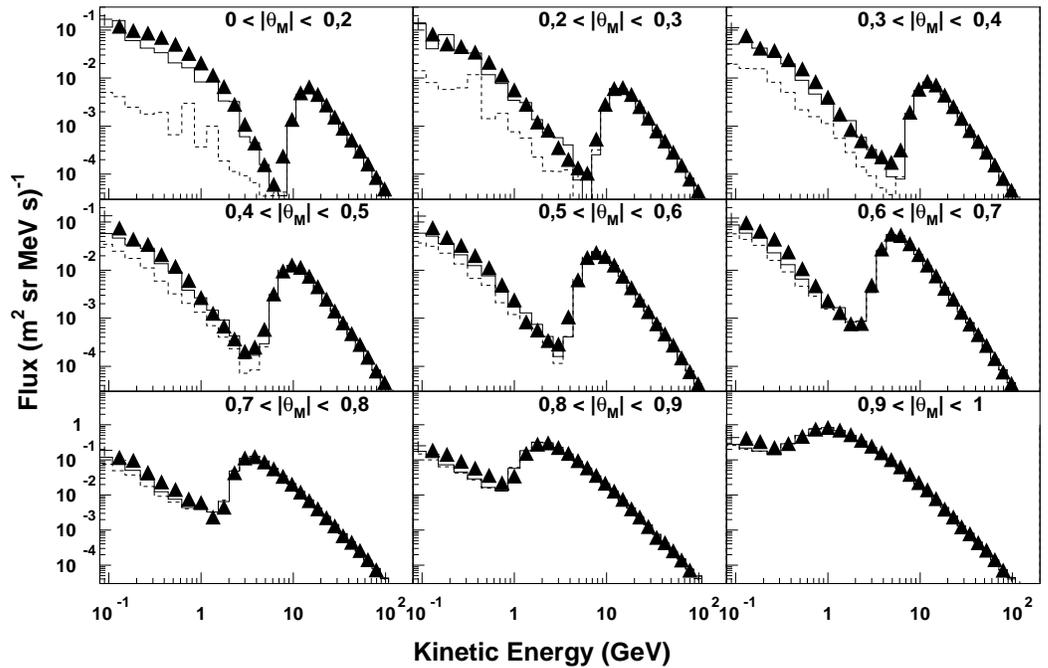,width=15cm}}
\caption{\label{subcut}
Proton spectrum measured~\protect\cite{AMS2}  at various geomagnetic latitudes
by the AMS detector on Space Shuttle compared to calculations
of Ref.~\protect\cite{Zuccon}.  The figure is from Ref.~\protect\cite{Zuccon}.
}
\end{figure}
%%%%%%%%%%%%%%%%%%%%%%%%%%%%%%%%%%%%%%%%

Several authors~\cite{Derome,Lipari2,Zuccon} 
have analyzed this ``second spectrum'' recently by performing
3-dimensional calculations which reproduce the data and allow them to make
conclusions about its origin and properties.  Since only a small fraction
of the secondary nucleons move away from Earth, the intensity of
the sub-cutoff population is significantly lower than would be the case
in the absence of a geomagnetic field.  In addition, a fraction of the
particles in the equatorial region remains trapped for several cycles,
so that the rate at which these particles re-enter the atmosphere is
lower (by a factor equal to the inverse of the number of trapped cycles)
than what is measured at 380~km.  The reduction of the re-entrant flux
is greatest in the equatorial region where the cutoff is generally high.
On average, the re-entrant sub-cutoff flux above the threshold
for significant pion production is always at least an
 order of magnitude below the flux at the cutoff energy~\cite{Zuccon}.
All these points are nicely
illustrated in the figure from Ref.~\cite{Zuccon}
that compares AMS data~\cite{AMS2} with their calculations.
The panels in Fig.~\ref{subcut}
show a succession of bins of geomagnetic latitude from
the equator ($\Theta_M=0$) to the poles ($\Theta_{M}=\pm1$).  The dashed
line shows the reduced rate at which the sub-cutoff particles re-enter the
atmosphere and so become available for neutrino production. 
In addition, the neutrino yield per incident particle is lower
below the cutoff than above.  As a consequence, the contribution
of the sub-cutoff particles to the neutrino flux is small, though
it remains to be calculated precisely.  

It is interesting to note
that the contribution of the subcutoff particles to the neutrino flux
is included in the one-dimensional calculation since all
secondary nucleons are assumed to go forward.

%\clearpage

\section{Hadronic interactions}

The third factor in Eq.~\ref{nuflux} is the yield of neutrinos
per incident cosmic-ray nucleon.  
For fixed $E_\nu$ the yields grow
strongly with increasing primary energy.  
At the same time, the primary spectrum
falls quickly as energy increases, to give distributions 
like those shown in Fig.~\ref{response1}.

Yields are calculated for events generated from primary cosmic rays
of various energies and mass, and the events are then weighted according
to an appropriate energy spectrum and mass composition.
Several technical details of the cascade calculations, in particular
muon energy loss and decay, require care
in their implementation.  The only fundamental source of uncertainty
in the yields, however, is from the treatment of particle production in
the individual collisions that make up the atmospheric cascades.

Two independent flux calculations~\cite{AGLS,Honda} 
have been widely used for 
interpretation of measurements of atmospheric neutrinos.  
Both are one-dimensional.
More recently, new, three-dimensional calculations have been 
made~\cite{Battistoni,Lipari3D,Hondanew}, 
which remove the major technical approximation
of the original calculations.  Differences among the calculations are
at the level of 20\% in overall magnitude and 5\% or less
in the ratios $\nu_e/\nu_\mu$ and $\bar{\nu}/\nu$.  
Although there are significant 
differences between the 3D and 1D 
calculations~\cite{Battistoni,Lipari3D,Hondanew},
especially in the angular distributions of sub-GeV neutrinos near the 
horizon, the major differences are due rather to differences in the treatment
of pion production and in the primary spectrum.  

\subsection{Cross sections}

%%%%%%%%%%%%%%%%%%%%%%%%%%%%%%%%%%%%%%%%
\begin{figure}[!htb]%14
\centerline{\psfig{figure=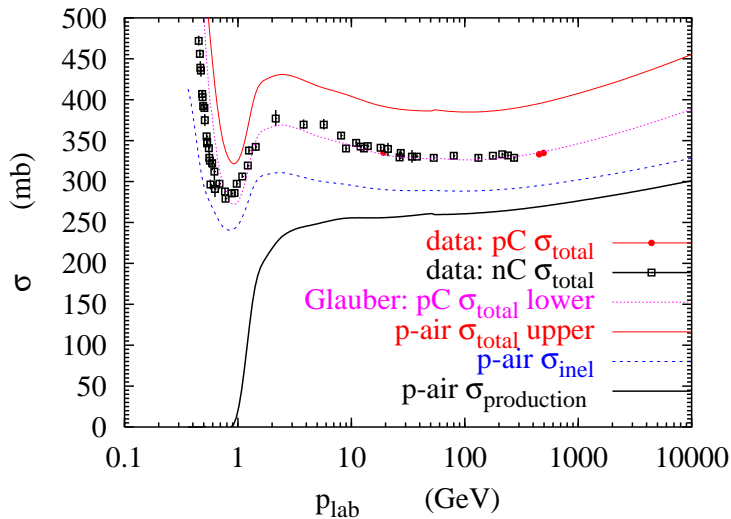,width=10cm}}
\caption{\label{sigma}
Plot of the total, inelastic and production p-air cross
sections.  Also shown is a comparison of the p-carbon
total cross section with accelerator data.  
The step in the p-air production
cross section at 1 GeV/c corresponds to the onset of 
significant pion production.
}
\end{figure}
%%%%%%%%%%%%%%%%%%%%%%%%%%%%%%%%%%%%%%%%

The starting point for calculation of the atmospheric cascade is a set of
energy-dependent interaction lengths for the various particles in
the cascade.  In the approximation that the target in each
interaction is treated as an average ``air" nucleus,
$\lambda_i({\rm g/cm}^2)\,\approx\, 14.5 m_p/\sigma$.
It is straightforward to treat the components separately with
partial interaction lengths for 78\% nitrogen, 21\% oxygen and 1\% argon
by volume.  If (as in Ref.~\cite{AGLS}) quasi-elastic interactions in which
the target nucleus fragments without pion production are not included when
nucleons interact in the atmosphere, then the
corresponding partial inelastic
cross section in which at least one secondary meson is produced
is used to obtain the interaction lengths.  If quasi-elastic
processes are included in the event generator (as in Ref.~\cite{FLUKA}) 
then the total
inelastic cross sections are used to obtain the interaction lengths.
Fig.~\ref{sigma} shows a plot of the nucleon-air cross sections
($\sigma_{tot}$, $\sigma_{prod}$ and $\sigma_{inel}$).   
The quasi-elastic cross section is
$\sigma_{QE}\;=\;\sigma_{inel}\,-\,\sigma_{prod}$.

The p-air cross sections have been calculated~\cite{Private} 
from the parameters
of the measured proton-proton cross section (elastic and total
cross section, slope parameter and ratio of real to imaginary
part of the forward scattering amplitude) using the Glauber multiple
scattering formalism~\cite{GM}.  To illustrate how well the formalism
works, Fig.~\ref{sigma} also shows the total cross section for carbon
calculated in the same way compared with an extensive collection of 
data~\cite{Dersch}. 
For lab energies corresponding to $\sqrt{s}>2$~TeV, where there are
not yet any direct measurements of the proton-proton cross sections,
the proton-air cross section becomes somewhat uncertain.  Since 
$\sqrt{s}=2$~TeV is equivalent to a lab energy of $\approx 2$~PeV,
this is not an important source of uncertainty even for 
the neutrino-induced muons.

Concerning treatment of incident nuclei, there are two aspects
to consider.  First, one needs a set of cross sections and interaction
lengths for the various nuclei.  In addition, an algorithm for
production of secondary particles is needed.
The usual approach is to neglect possible coherent
effects and use the Glauber multiple scattering theory
to determine a number of nucleon-nucleon interactions in each encounter.
This can be done with a monte carlo realization of the multiple
scattering calculation to determine the number of interacting pairs
event-by-event.  Then
the model for pion and kaon production in nucleon-nucleon
collisions is invoked.  A further approximation sometimes used is
to determine the number of wounded nucleons in the projectile and
then immediately use the model for particle production in nucleon-air 
collisions.  In this approximation~\cite{JEngel}, 
incident nuclei of mass A and 
energy $E$ are equivalent to A nucleons each of energy per
nucleon $E/A$.

\clearpage

%%%%%%%%%%%%%%%%%%%%%%%%%%%%%%%%%%%%%%%%
\begin{figure}[!htb]%15
\centerline{\psfig{figure=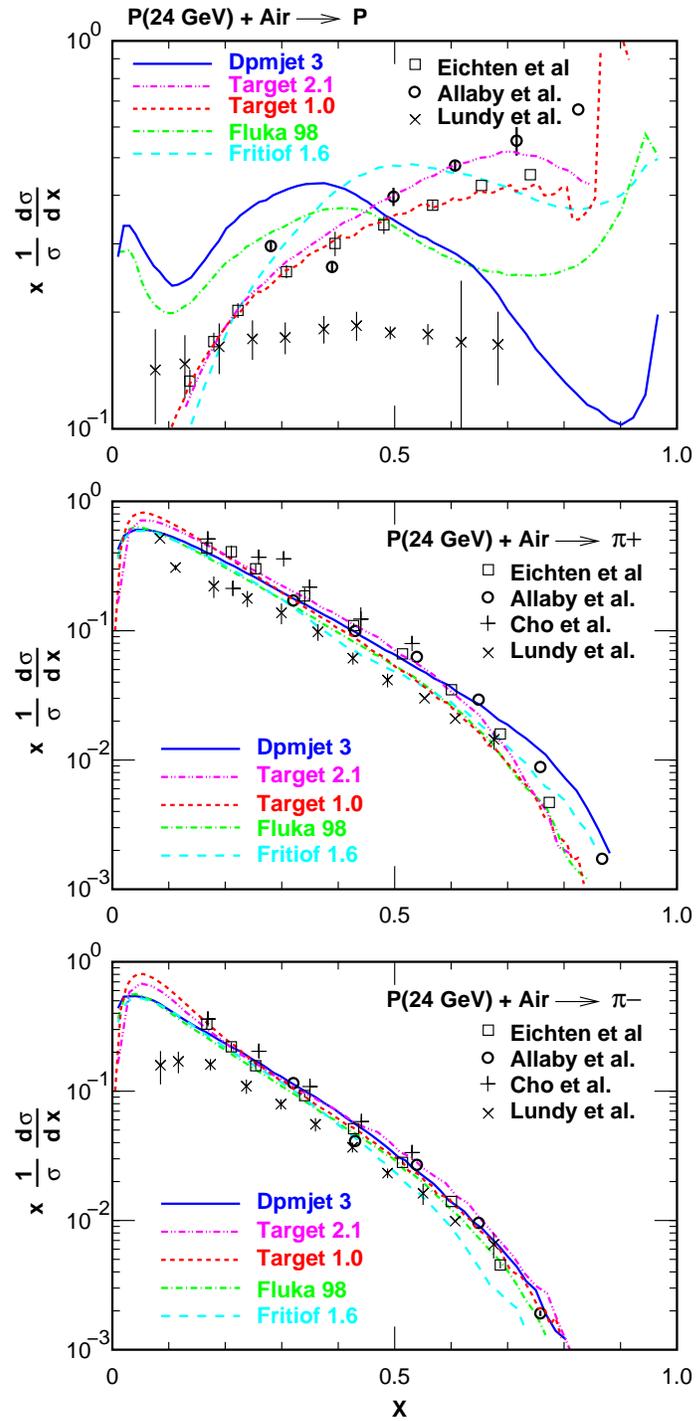,width=9cm}}
\caption{\label{hondaplot}
Inclusive cross sections for protons (top); $\pi^+$
and $\pi^-$ (bottom).
}
\end{figure}
%%%%%%%%%%%%%%%%%%%%%%%%%%%%%%%%%%%%%%%%

\subsection{Pion and kaon production by protons}

The models of hadroproduction in use for calculations
of the atmospheric neutrino
flux are based in one way or another on accelerator data for protons on
light nuclei.  Most such experiments were done as studies for
design of accelerator neutrino beams.  A good example is the work of
Cho {\it et al.} using 12.4 GeV/c beams at the Argonne ZGS
on a beryllium target~\cite{Cho}.  They measured $d\sigma/dpd\Omega$
as a function of secondary particle momentum with the spectrometer
at several angles.  They then used a standard 
four-parameter fit to the double differential
cross section~\cite{SW} to find smooth fits to the data.
Similar experiments in the region around 20~GeV/c
were made at the CERN~PS~\cite{Allaby,Eichten} and at 
15~GeV/c at BNL~\cite{Abbott}.
Figs.~\ref{hondaplot} %,\ref{hondaplot2} and~\ref{hondaplot3} 
%shows the data from these and other~\cite{Dekker,Lundy} experiments
shows the data from these and other~\cite{Lundy} experiments
as inclusive distributions integrated over angle using the
same parameterization~\cite{SW} to interpolate and extrapolate 
into unmeasured regions of phase space.  The integration requires
an extrapolation into
unmeasured regions of transverse phase space, which leads to
uncertainties in addition to those associated with the measurements,
as discussed in Ref.~\cite{Engel}.

%%%%%%%%%%%%%%%%%%%%%%%%%%%%%%%%%%%%%%%%
\begin{figure}[!htb]%16
\centerline{\psfig{figure=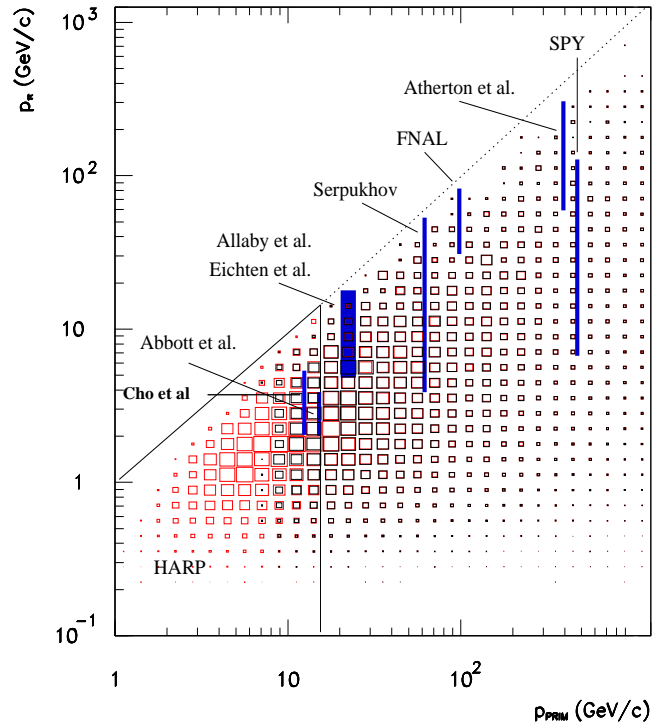,width=9cm}}
\caption{\label{Raeven}
Weighted phase space distribution for sub-GeV
atmospheric neutrinos.
}
\end{figure}
%%%%%%%%%%%%%%%%%%%%%%%%%%%%%%%%%%%%%%%%

The coverage of phase space in the accelerator measurements 
is generally not matched ideally to the atmospheric neutrino problem.
Fig.~\ref{Raeven}
shows the regions of phase space for proton-air interactions that
contribute to the sub-GeV neutrino signal at Super-K, as calculated with
the FLUKA program~\cite{FLUKA,BatPrivate}.  
The important region shifts to the right for higher energy
neutrinos in a way that can be estimated from the response functions
discussed above (see Fig.~\ref{response1}).
Superimposed on the figure
are the regions of phase space covered by the main data sets used
for tuning models of hadroproduction for calculation
of atmospheric neutrinos.  One can note in particular that the
region for $p_{beam}\ge20$~GeV/c and $p_\pi<4$~GeV that is
important for the sub-GeV events is not covered.

The curves in 
Fig.~\ref{hondaplot} % Fig.~\ref{hondaplot2} and Fig.~\ref{hondaplot3} 
show some of the inclusive cross sections 
in use for calculations of atmospheric neutrinos.  
Dpmjet3~\cite{Dpmjet3} is the event generator currently 
used by Honda et al.~\cite{Hondanew}, who previously~\cite{Honda} used
Fritiof~1.6~\cite{Fritiof1-6}.  Target~1.0 is the event generator
of the Bartol group~\cite{AGLS}, and Target~2.1 is a preliminary
revised version~\cite{Hamburg}.  The calculation of Battistoni
et al.~\cite{Battistoni} uses the interaction model 
embedded in the FLUKA~\cite{FLUKA} cascade program. 

The differences
shown here are one of two main
components responsible for the differences at the
level of 15\% among the calculations of Refs.~\cite{AGLS,Honda,Battistoni}
and~\cite{Hondanew,Hamburg}.  The other main factor is the treatment of
the primary spectrum.  We comment further on these differences
in the next section.

\begin{figure}[!thb]%17
\centerline{{\psfig{figure=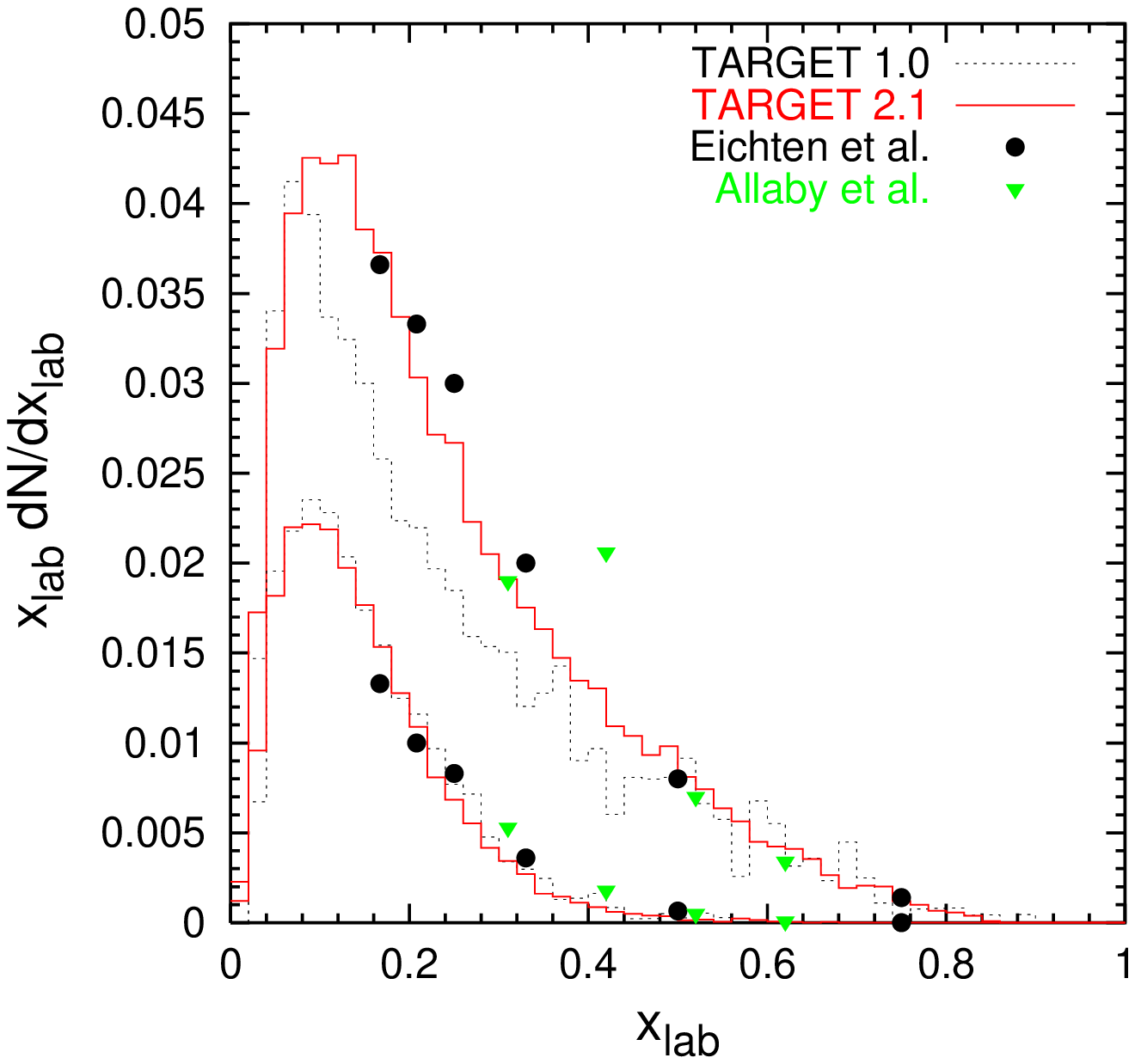,width=6.cm} \hspace{5mm}
{\psfig{figure=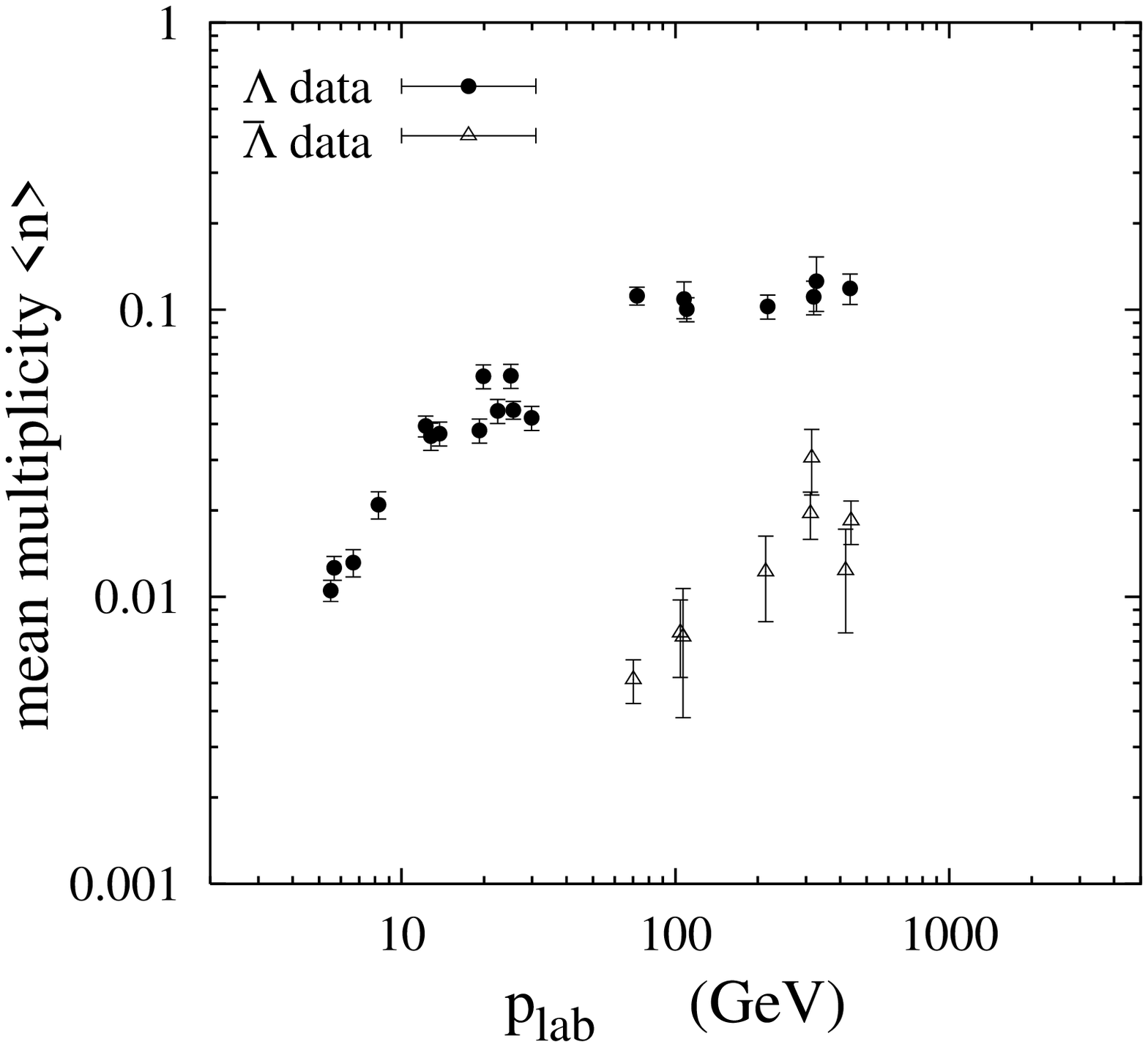,width=6.cm}}}}
\caption{
Left panel: Integrated inclusive cross sections for
kaon production from Ref.~\protect\cite{Engel}.
The upper set of points and histograms is for $K^+$,
while the lower is for $K^-$.
Right panel: Multiplicity of $\Lambda$ and $\bar{\Lambda}$
production in proton-proton interactions 
from Refs.~\protect\cite{Lam}
}
\label{strange}
\end{figure}

The process $p\rightarrow\Lambda K^+$ makes an important
contribution to production of strangeness, and it accounts for the
large excess of $K^+$ over $K^-$ in the fragmentation region.
The effect is clearly visible in Fig.~\ref{strange}a, and it becomes
increasingly important at high energy as kaons begin to
dominate the neutrino flux (see Fig.~\ref{PiK2}.)  Fig.~\ref{strange}b
summarizes data on production of $\Lambda$ and $\bar{\Lambda}$ in 
pp collisions.  The difference ($\Lambda\,-\,\bar{\Lambda}$) corresponds
to $p\rightarrow\Lambda K^+$.  
Because of the steep primary spectrum, this process is weighted
heavily in the spectrum-weighted moments for kaons
that enter into the calculated fluxes of neutrinos.

\subsection{High energies}

At sufficiently high energy the most important contribution
to the neutrino flux will come from decay of charmed hadrons, the 
``prompt'' flux, so called because of the short lifetimes of
charmed particles.  
The critical energy for charm decay is $\sim 5\times 10^7$~GeV.
Thus the neutrinos from decay of charmed hadrons
continue with the same spectral index as the primary
cosmic-ray spectrum up to this energy, while the neutrinos
from decay of pions and kaons become steeper at much lower
energy.  Moreover, the prompt contribution is isotropic up to
$\sim 10^7$ GeV while the contribution from pions and kaons
is proportional to $\sec(\theta)$ for $E_\nu\gg 1$~TeV.
As a consequence, even though $Z_{\rm charm}\ll Z_{\pi,K}$,
the contribution from charm decay dominates
the neutrino spectrum at sufficiently high energy.

Predictions in the literature for the crossover energy
differ by several orders of magnitude. Costa~\cite{Costa}
surveys a wide range of calculations of prompt leptons.
An analysis of the angular dependence of measurements of high-energy
atmospheric muons
(fitting to an isotropic plus a secant~$\theta$ term)~\cite{VolkovaC}
suggests that the crossover energy may be as low as $\sim10$~TeV.
At the other extreme, in a perturbative QCD calculation, 
Thunman et al.~\cite{Gondolo} showed the crossover only
around $E_\nu\sim 1000$~TeV.  A more recent improved QCD
calculation~\cite{Reno} puts the crossover around $~100$~TeV.

A concern with the perturbative QCD approach in the context of prompt
cosmic-ray leptons is that it does not explicitly include
the charm analog of $p\rightarrow\Lambda K^+$ discussed above.
The $\Lambda$ multiplicity plot in Fig.~\ref{strange}b allows
an estimate of the probability per interaction of producing
a forward $\Lambda K^+$ pair as $\approx 0.04$ (by subtracting
$\bar{\Lambda}$ from $\Lambda$ and dividing by 2 to get the 
forward-moving particles).  If we assume the production 
of massive hadrons with similar quark content 
is inversely proportional to the square of the masses~\cite{GH}, then
we can estimate the probability  of the corresponding
charmed process $p\rightarrow\Lambda_C D^0$ as
\begin{equation}
\label{charm1}
P_{\rm forward\;charm}\;\sim\;0.04\times\left(M_\Lambda + M_K\over
M_{\Lambda_C}+ M_D\right)^2\;\sim\;0.006.
\end{equation}

The actual calculation of the prompt lepton spectrum requires
an assumption about the shapes of the inclusive cross section
for production of the various charmed hadrons as well as
accounting for the various relevant branching ratios and decay
distributions. Bugaev {\it et al.}~\cite{Bugaev} have systematically
calculated prompt muon production in a recombination quark-parton
model, which explicitly includes the important fragmentation process.

\section{Comparison of neutrino flux calculations}

%%%%%%%%%%%%%%%%%%%%%%%%%%%%%%%%%%%%%%%%
\begin{figure}[!htb]%18
\centerline{\psfig{figure=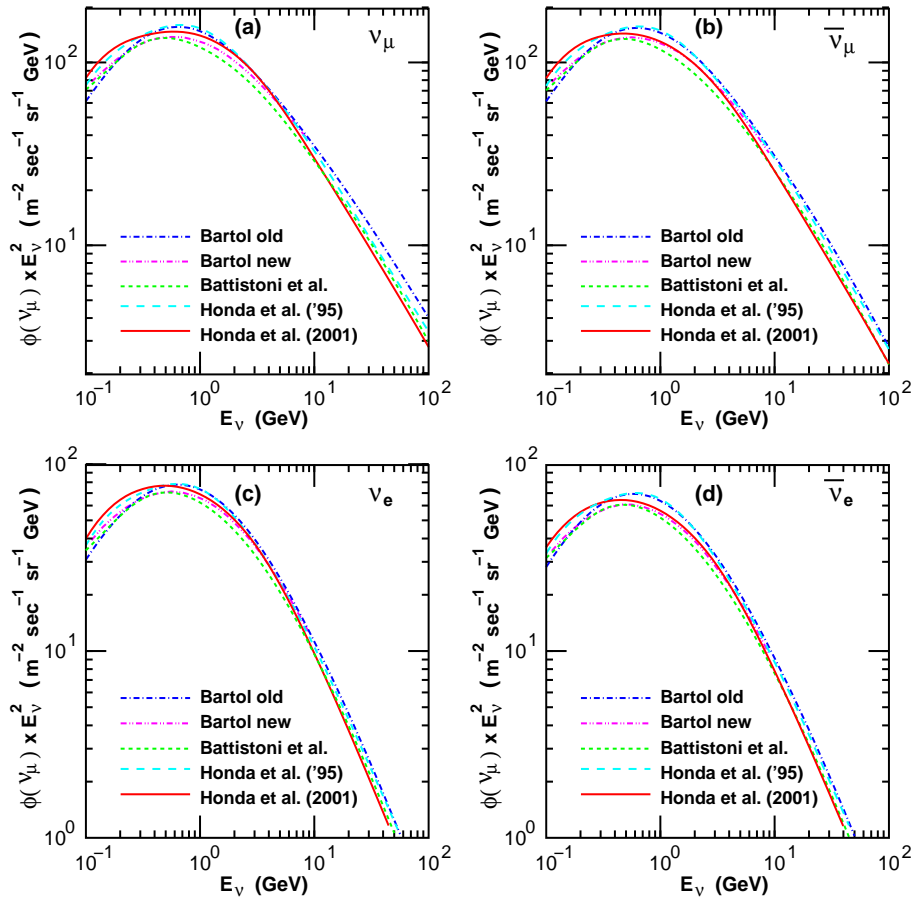,width=12cm}}
\caption{\label{compnu}
Comparison of neutrino flux calculations for the
location of Kamioka averaged over all directions.
}
\end{figure}
%%%%%%%%%%%%%%%%%%%%%%%%%%%%%%%%%%%%%%%
Comparing independent calculations is one way to assess
the uncertainty in our knowledge of the atmospheric neutrino flux.
Fig.~\ref{compnu} is a plot of the flux at Kamioka as calculated by
three groups.  The calculations of the Bartol group~\cite{AGLS} and
Honda {\it et al.}~\cite{Honda}, both one-dimensional, are quite
similar between 300 MeV and 10 GeV, but the similarity results to
some extent from compensating differences in input.  The primary
spectrum normalization is higher in Ref.~\cite{Honda} than in Ref.~\cite{AGLS}
while the pion production is lower.  In contrast, the three-dimensional
calculation of Battistoni {\it et al.}~\cite{Battistoni} 
uses exactly the same primary
spectrum as Ref.~\cite{AGLS} and finds a neutrino flux some 15\%
lower in the same energy region.  As mentioned above, this 
difference is primarily a consequence of  differences in pion production
and not a consequence of the 3-dimensional nature of the calculation.
The new one-dimensional calculation of Honda {\it et al.} uses a new
model for pion production and a new and lower primary spectrum.

The interaction models used by these three groups represent three
different approaches to the problem.
Target is a simple phenomenological representation of pion and kaon production
in interactions of protons, pions and kaons with light nuclei~\cite{Hamburg}.
Energy conservation is used as an important constraint in adjusting
fits to the data.   When events are generated, the fragment nucleon
from the projectile is selected first, and the remaining
energy distributed among pions.  As a consequence, the fit to
the fast nucleons is of primary importance.  
Honda et al. use more sophisticated event generators designed
for monte carlo studies at accelerators, Fritiof 1.6 in Ref.~\cite{Honda}
and Dpmjet3 more recently~\cite{Hondanew}. 
FLUKA~\cite{FLUKA} is a cascade code that includes generation and propagation 
of secondary particles
of all types in hadronic interactions of all types as an integral
part of a cascade code adaptable for any purpose.

\begin{figure}[!thb]%19
\centerline{{\psfig{figure=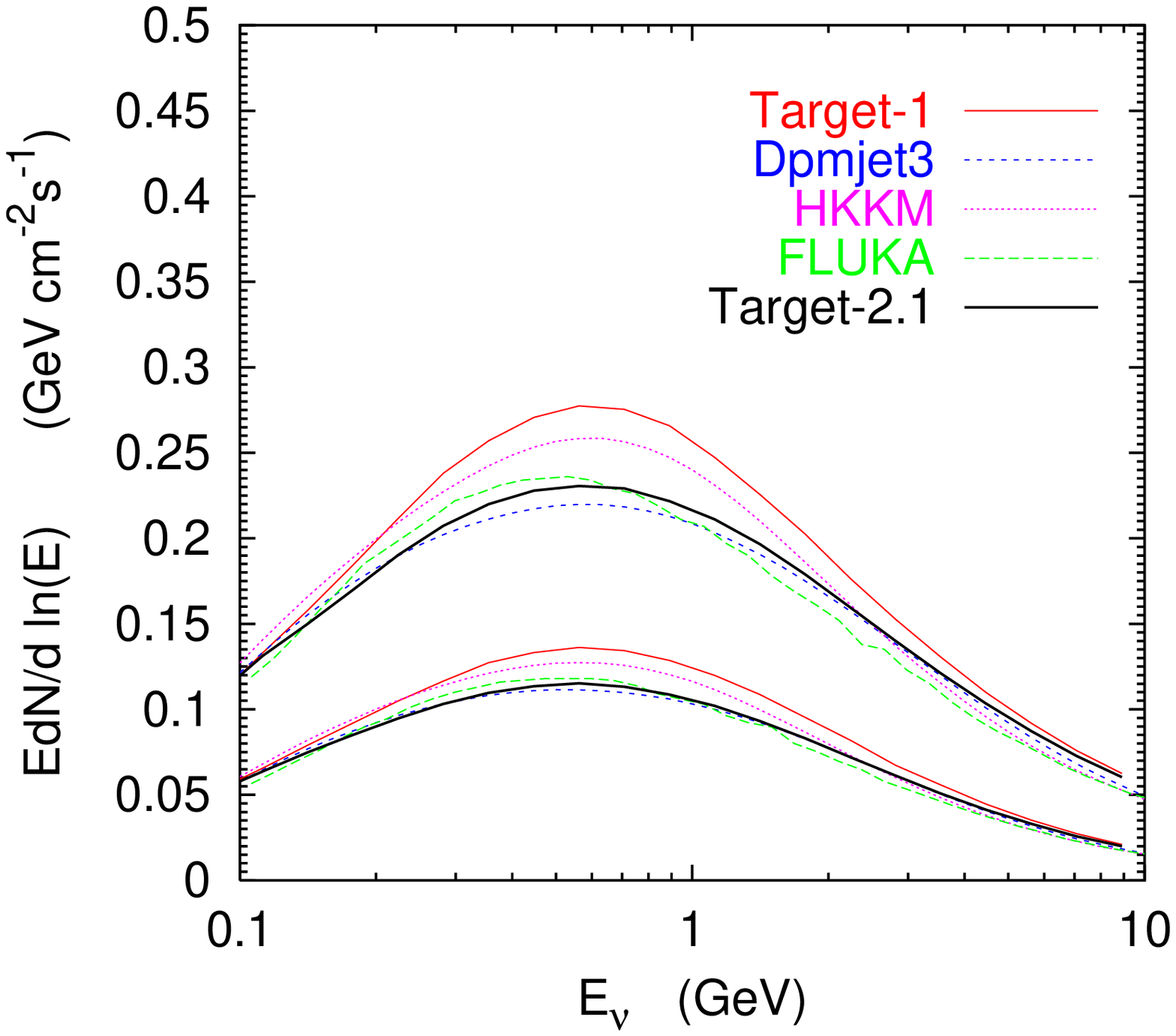,width=6.4cm} \hspace{1mm}
{\psfig{figure=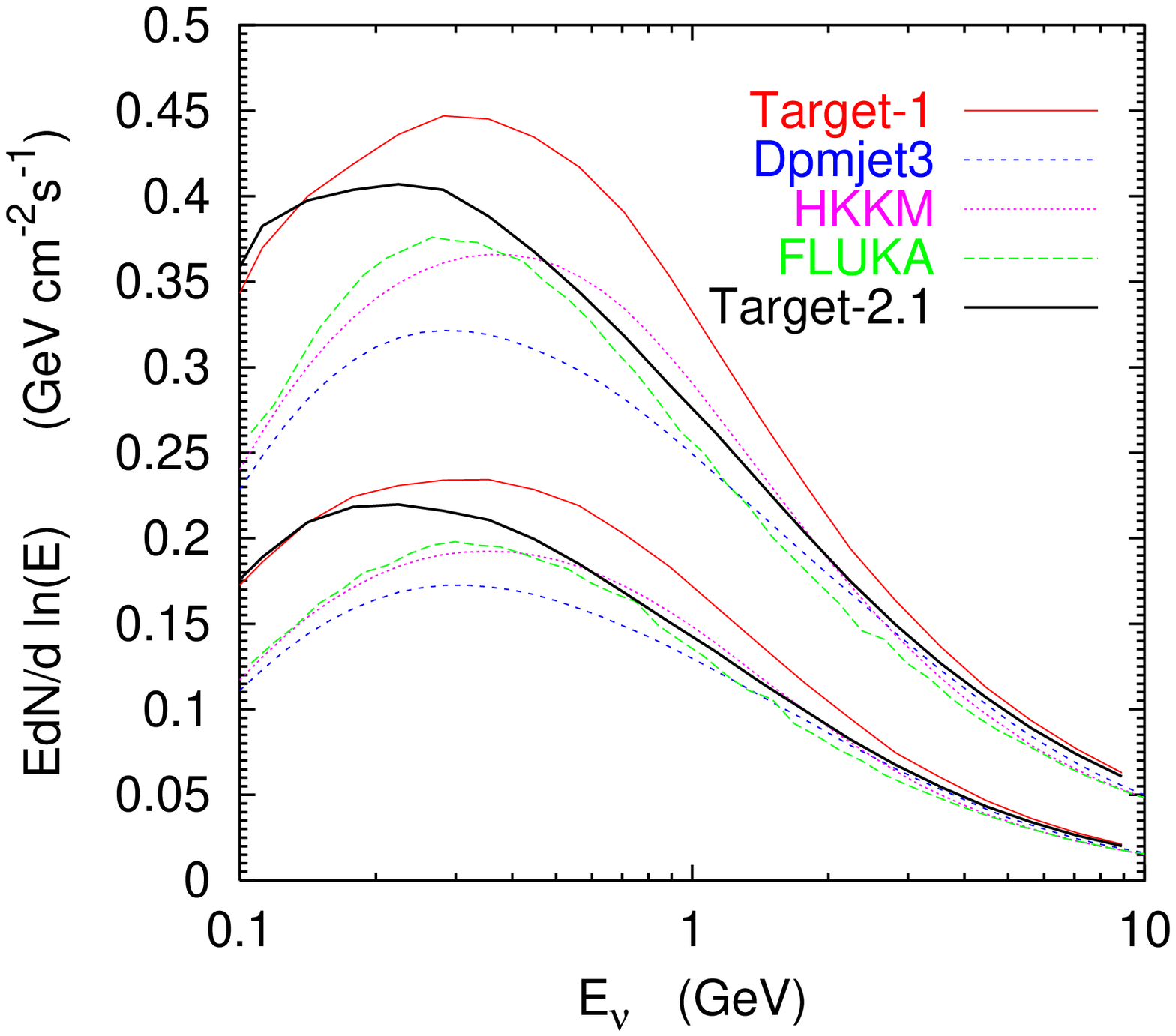,width=6.4cm}}}}
\caption{
Angle-integrated neutrino fluxes ($\nu+\bar{\nu}/3$)
at Kamioka (left) and Soudan (right)
with several different interaction models:
HKKM~\protect\cite{Honda}, Dpmjet3~\protect\cite{Hondanew},
FLUKA~\protect\cite{Battistoni}, 
Target~1.0~\protect\cite{AGLS} and Target~2.1~\protect\cite{Hamburg}.
The upper set of curves is for muon neutrinos and the lower set
for electron neutrinos.
}
\label{tar-tom}
\end{figure}

To focus on differences caused by the treatment of hadronic interactions,
we refer to Fig.~\ref{tar-tom}.
Exactly the same primary spectrum and composition (that from Ref.~\cite{AGLS})
have been used by Battistoni et al.~\cite{Battistoni} and by the Bartol
group~\cite{AGLS,Hamburg}.  Although different
primary spectra have been used in Refs.~\cite{Honda,Hondanew}, the results
shown here are from special calculations with the hadronic interaction
models used in the published paper but using a primary spectrum identical
to that of Ref.~\cite{AGLS}.
Therefore, the only differences among these results
are due to the differences in treatment of hadronic interactions.  
It should be possible to trace the differences in the neutrino fluxes shown here
back to differences in interaction models such as those shown in Fig.~\ref{hondaplot}.
For example, the excess of pions in $\sim20$~GeV interactions in Target~1.0
as compared to Target~2.1 and FLUKA corresponds to the larger calculated
flux at Kamioka.  The larger difference among the three calculations at Soudan points to differences in treatment of low-energy
interactions that are important at Soudan but not at Kamioka with its higher cutoff.

In the past two years several new calculations of 
the flux of atmospheric neutrinos
have appeared.  They are summarized in Table~\ref{Tab3} 
with a comment indicating
their normalization relative to Ref.~\cite{Battistoni}.
Some of the recent calculations show rather large differences from those
discussed so far.  
The comparisons discussed above, however, suggest that
uncertainties in normalization due to hadronic interactions are  $<\pm15$\% 
(and somewhat larger at higher energy where the kaon contribution 
is more important).
Differences in ratios such as $\nu_e/\nu_\mu$ are significantly
smaller, at the level of $5$\% (see Fig.~\ref{nuratio}).  
Combining a $\pm15$\% estimate of
uncertainty from hadronic interactions in quadrature 
with an estimate of $\pm20$\%
for the primary spectrum gives an estimated uncertainty of $\pm25$\%.
Even if the estimates are simply added algebraically, it seems difficult
to account for differences of a factor of two within the uncertainties
in the input.

\begin{table}[!htb]
\def~{\hphantom{0}}
\caption{Calculations of the atmospheric neutrino flux}
\begin{tabular}{@{}|r|r|l|@{}} %\hline
\toprule
Reference &  & Comment  	\\ %\hline
\colrule
FLUKA~\cite{Battistoni} & 3D & 1 \\ \hline
New Bartol~\cite{Hamburg} & 1D & $\approx1$ (preliminary) 	\\ \hline
Honda {\it et al.}~\cite{Hondanew} & 3D & $+10$ to $15$\%  \\ \hline
SNO sub-group~\cite{Tserk} & 3D & $\sim1$  \\ \hline
Fiorentini {\it et al.}~\cite{Fior} & 1D & $-20$\% for sub-GeV; $\sim1$  for $>$GeV\\ \hline
CORSIKA~\cite{Wentz} & 3D & $\sim1$ (preliminary)  \\ \hline
Grenoble~\cite{Liu} & 3D & $\sim {1/2}$  \\ \hline
Plyaskin~\cite{Plyaskin} & 3D & $< {1/2}$  \\ %\hline
\botrule
\end{tabular}
\label{Tab3}
\end{table}

\section{Three-dimensional calculations}

The major technical limitation of most calculations of the atmospheric
neutrino flux until recently is that they are one-dimensional; in other
words, it is assumed that all secondaries move in the direction of
the primary particle from which they descend.  This approximation neglects
the transverse momentum of the secondaries as well as their bending in the
geomagnetic field as they propagate.  The latter is most important for muons 
and their decay products because of the muon's long decay length.  
A muon with
Lorentz factor $\gamma$ typically bends by an angle
\begin{equation}
\label{bending}
\theta\;\sim\;\gamma\,c\,\tau_\mu\,/\,r_L,
\end{equation}
where $r_L\,\approx\,E_\mu\,/\,eB$ is its gyroradius in the geomagnetic field.
Dependence on $E_\mu$ cancels in Eq.~\ref{bending}, 
and typical bending is $\sim 3^\circ$ for a muon
in a field of order $0.3$~Gauss.  This is independent of energy for particles
with trajectories such that their potential pathlength before reaching the ground
exceeds their decay length.  As a consequence, muon bending can be noticeable
even in the multi-GeV range, as in the case of the modification of the East-West
effect reviewed in \S4.2.

The effect of transverse momentum, on the other hand, becomes increasingly important at
low energy.  Since the distribution of transverse momentum of secondaries in hadronic
interactions is nearly independent of energy, with
$\langle p_T\rangle\,\approx\,300$~MeV for pions, 
the corresponding angular deviation is inversely proportional to energy, of order
\begin{equation}
\label{ptran}
{p_T\over E_\pi}\;\sim \;{0.1\over E_\nu({\rm GeV})}
\end{equation} 
in radians.  The 3D effects are thus most important in the sub-GeV region. 

Three-dimensional calculations of the neutrino flux present a significant
technical challenge.  A full 3D calculation without any approximations has
to sample primary cosmic rays uniformly and isotropically in the downward
half hemisphere at every point on the globe, taking account of the cutoff
for every direction at each point.  The created neutrinos will then also
be nearly isotropic in the downward hemisphere (except for geomagnetic cutoff
effects), and all but a small fraction of order $A/R_\oplus^2\sim 10^{-5}$ 
will be discarded.  (Here $A$ is the projected area of the detector
and $R_\oplus$ the radius of the earth.)
On the other hand, in a one-dimensional calculation, 
only cosmic rays heading toward the detector are sampled, and 
all the neutrinos go through the detector.
Thus, the one-dimensional calculation is a very efficient approximation.

Several approaches have been used to make the 3-dimensional 
calculation more tractable.  
Tserkovnyak et al. assumed a huge detector size\cite{Tserk};
Battistoni et al. assumed spherical symmetry, 
ignoring the geomagnetic field inside 
the atmosphere\cite{Battistoni}; and
Honda et al. assumed a dipole magnetic field and also a huge 
detector\cite{Honda3D}. 
In his 3D calculation~\cite{Lipari3D} Lipari divided
the Earth's surface into five 
detector zones.  
A preliminary version of a full 3D calculation without
approximations uses a modified 
version of Corsika and the DPMjet interaction model~\cite{Wentz}.

\begin{figure}[!htb]%20
\centerline{\psfig{figure=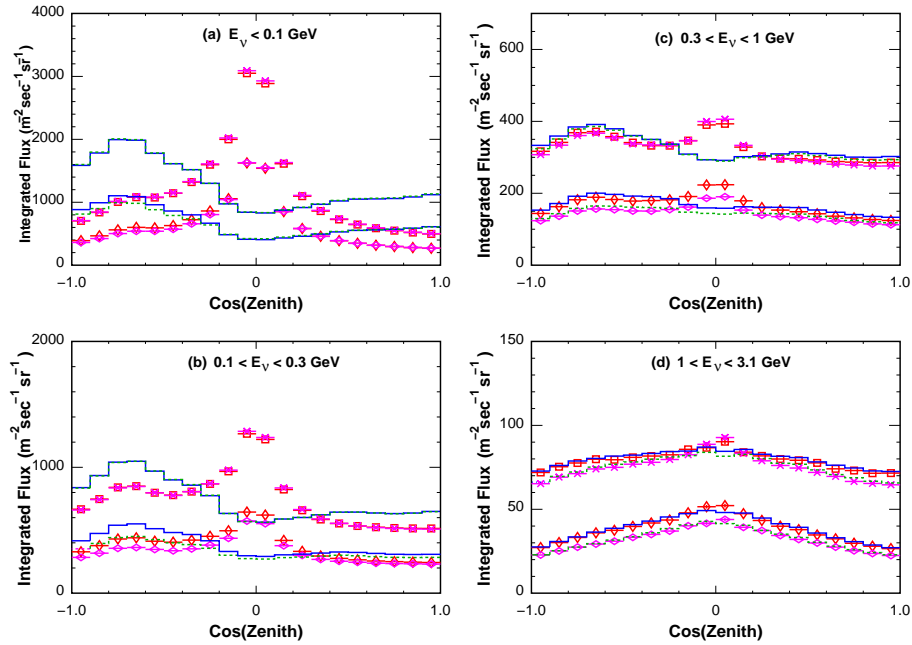,width=12.cm}}
\caption{Comparison of three-dimensional and 
one-dimensional calculations of neutrino fluxes
at Super-K~\cite{Honda3D}.
Squares are for $\nu_\mu$, asterisks for $\bar \nu_\mu$, vertical diamonds
for $\nu_e$, and horizontal diamonds for $\bar \nu_e$ for three dimensional
calculation.  The solid line histograms show the $\nu$ fluxes, 
and the dotted ones the $\bar\nu$ fluxes for 
the one dimensional calculation.  The panels show four
bins of neutrino energy: a) 0~--0.1~GeV,
b) 0.1~--~0.3~GeV,
c) 0.3~--~1~GeV, and d) 1~--~3.1~GeV.  The fluxes are averaged 
over all azimuthal angles in 20 bands of $\cos(\theta)$ from
$-1$ (straight up) to $+1$ (vertically downward moving).
}
\label{Fig-3d-1d}
\end{figure}

\subsection{Geometry of neutrino production}

The most prominent feature of three-dimensional calculations is
the neutrino flux enhancement near the horizon, as illustrated
in Fig.~\ref{Fig-3d-1d} from Ref.~\cite{Honda3D}.
This enhancement is a geometrical effect not present in the one-dimensional
calculations.  Fig.~\ref{Fig-3d-1d} compares 3D and 1D calculations
specifically at Kamioka, where the angular dependence results from
a complex combination of the physics of neutrino production, geomagnetic effects
and geometry.  Before discussing the figure in detail, a digression
on the geometry of neutrino production will be helpful.

\begin{figure}[!thb]%21
\centerline{\psfig{figure=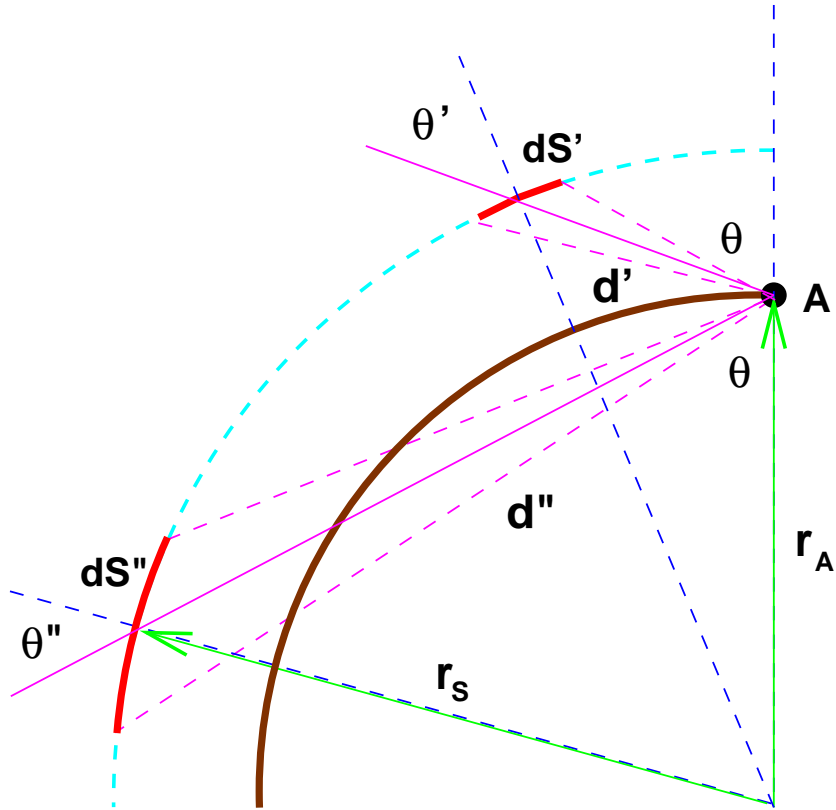,width=11.cm}}
\caption{Geometry of atmospheric neutrino production (see text).
The detector is at A, and we compare a neutrino entering from
above the horizon at zenith angle $\theta$ with one entering
from below the horizon at the same nadir angle.  The angles
$\theta^\prime = \theta^{\prime\prime}$ are the apparent local
zenith angles in the production regions, indicated by the heavily
shaded arcs.
}
\label{earth}
\end{figure}

Lipari~\cite{Lipari3D} has given a detailed explanation of the
origin of the excess of low energy neutrinos near the horizon
in the 3D calculation, starting from an analysis in which all
neutrinos are produced in a shell of constant altitude $h\sim15$~km
and geomagnetic effects are neglected.  
We follow his argument, generalizing it somewhat to display also the
up-down symmetry of the neutrino flux that characterizes both the
3D and the 1D calculations in the absence of geomagnetic effects.
%%%%%%%%%%%%%%%%%%%%
%It is convenient to consider a spherical detector with radius $a$.
Fig.~\ref{earth} illustrates the
detector $A$ at a distance $r_A\approx R_\oplus$ from the center of
the earth  surrounded by a shell of radius 
$r_S\approx R_\oplus + h$ centered around the Earth.  
Eventually, 
an integral has to be made over the production altitudes ($h$) of
the neutrinos.  
Most neutrino 
production occurs in a band of altitude $10 < h < 20$~km,
which is a thin shell compared to the radius of the Earth. 

We consider neutrino emission from two patches of the 
production shell with area $dS'$ and $dS''$ above and below
the horizontal at the same zenith and nadir angles, $\theta$ and with the 
same solid angle at the detector, 
\begin{equation}
\label{up-down}
d\Omega=dS'\cos\theta'/(d')^2=
dS''\cos\theta''/(d'')^2. 
\end{equation}
The distance to the patch
above the horizon is $d^\prime =\ell$ and the distance to the 
patch below the horizon is $d^{\prime\prime}$, 
while the local zenith angles at the patches are $\theta^\prime$
and $\theta^{\prime\prime}$.  By a geometrical construction 
one can show that the angles are related by
\begin{equation}
\label{geom1}
\cos\theta^\prime \;=\;\cos\theta^{\prime\prime}
\;=\;\sqrt{1\,-\,(r_A/ r_S)^2\,\sin^2\theta}
\;\ge\;\sqrt{1\,-\,(r_A/ r_S)^2}\;\approx\;0.07.
\end{equation}

Next let
$Y_{\nu_i}(E_\nu,\Omega')\;dE_\nu\;dS'\;d\Omega'$
be the number of neutrinos of type $i$ emitted from the patch $dS'$
in energy interval [$E_\nu, E_\nu + dE$]
into the solid angle $d\Omega' = d\cos\theta'd\varphi'$ centered at
$\Omega' = (\cos\theta', \; \varphi')$, per unit time.
We represent the projected area of the detector as a
function of zenith angle as ${\cal A}(\theta)$.  Then, if the
detector is up-down symmetric
(i.e. ${\cal A}(\theta)={\cal A}(180^\circ -\theta)$), the solid
angle subtended by the detector is ${\cal A}/(d^\prime)^2$ above
the horizon and ${\cal A}/(d^{\prime\prime})^2$ below.  The
corresponding rates of neutrinos through the detector are
$Y_{\nu_i}(E_\nu,\Omega')\; dE_\nu\; dS' \times {\cal A}/(d')^2$
and
$Y_{\nu_i}(E_\nu,\Omega'')\; dE_\nu\; dS'' \times {\cal A}/(d'')^2$
respectively.
The differential rate from the patch above the horizon is
\begin{equation}
\label{geom2}
{dN_{\nu_i} \over dE_\nu d\Omega  d{\cal A}} =
{Y_{\nu_i}(E_\nu, \Omega')\; dE_\nu\; dS' \times ({\cal A}/(d')^2) \over
dE_\nu (dS'\cos\theta'/(d')^2) ({\cal A})}
= {1 \over\cos\theta'} Y_{\nu_i}(E_\nu, \Omega')
\end{equation}
with the same result for the patch below the horizon.
Eq.~\ref{up-down} has been used for $d\Omega$.
Since the distances cancel and $\theta^{\prime\prime} = \theta^\prime$,
the flux is symmetric from above and below the horizon no
matter how complicated the dependence of $Y_\nu$ on angle,
provided it is the same above and below the horizon (i.e. in the
absence of geomagnetic effects, mountains, seasonal variations, etc.).
To the extent that the detector response is also symmetric, the
measured rates will be as well.
The factor $1/\cos\theta^\prime$ in Eq.~\ref{geom2}
is the origin of the enhancement near the horizontal
in the 3D calculation.  It comes from the expression~\ref{up-down}
for the solid angle of the production region, and it is limited by
the inequality in Eq.~\ref{geom1}.

The relation between the yield for a patch of the production shell
and the incident cosmic-ray flux is
\begin{equation}
\label{3dyield}
%\begin{array}{ll}
Y_{\nu_i}(E_\nu, \Omega')\,= \,\int_{E_{cr}}\int_{\Omega_{cr}}\;\;[\eta_{\nu_i}(E_\nu,\Omega' :
E_{cr},\Omega_{cr})\,]
 \times \; [(dN_{cr} / dE_{cr}) \, \cos\theta_{cr}\,] \; d\Omega_{cr} d E_{cr},
%\end{array}
\end{equation}
where 
$\eta_{\nu_i}(E_\nu, \Omega': E_{cr},\Omega_{cr})\,$
is the differential probability with which 
primary cosmic rays with energy $E_{cr}$
and arrival direction $\Omega_{cr} = (\cos\theta_{cr}, \; \varphi_{cr})$
produce $\nu_i$ in the energy interval  
[$E_\nu, E_\nu + dE$] and solid angle $d\Omega'$.  
Apart from geomagnetic effects the flux
of cosmic rays, $dN_{cr}/ dE_{cr}$ 
$({\rm cm^2\cdot sr\cdot s \cdot GeV})^{-1}$,
is isotropic.  
The factor $\cos\theta_{cr}$ in Eq.~\ref{3dyield} is the projection
of the isotropic flux onto the surface element $dS^\prime$ that is
needed to obtain the rate at which primaries enter the production
shell from a particular arrival direction.

As energy increases neutrino production becomes peaked in the
forward direction ($\cos\theta_{cr} = \cos\theta^\prime$).
In the high energy limit
$\eta_{\nu_i}(E_\nu, \Omega')\rightarrow 
\eta_{\nu_i}(E_\nu, \cos\theta') \times \delta(\Omega', \Omega_{cr})$,
and we recover the one-dimensional approximation.
Then $Y_{\nu_i}(E_\nu, \Omega')\propto\cos\theta^\prime$
and the horizontal enhancement in Eq.~\ref{geom2} is exactly canceled.
For low energy neutrinos the deviation from the direction of
the parent primary cannot be neglected and the cancellation is
not complete.  Eq.~\ref{ptran} sets the energy scale 
for the horizontal enhancement.

We return now to a discussion 
of the results displayed in Fig.~\ref{Fig-3d-1d}.
The intrinsic angular dependence of the one-dimensional calculation 
in the absence of the geomagnetic field is displayed
in the left panel Fig.~\ref{nuratio} in \S2.1.  
When geomagnetic effects are unimportant, there is an excess
of neutrinos near the horizon.  At Super-K, however, there are regions of
very high cutoff rigidity near the horizon, which leads to the suppression
of low-energy neutrinos from near $\cos\theta=0$ for the 1D calculation
shown by the line histograms in Fig.~\ref{Fig-3d-1d}.  The characteristic
peaking near the horizon only appears again at Super-K for $E_\nu > 1$~GeV
(panel d in the figure).  At this energy the three-dimensional effects are
also negligible, and the 1D and 3D calculations agree.  At lower energy,
however, the flat behavior or suppression near $\cos\theta\sim 0$
is replaced with the horizontal peak characteristic of the correct 3D
treatment.  

The horizontal enhancement of the neutrino flux appears mainly at low energy
and is therefore essentially impossible to detect because of the large angle
between a low energy neutrino and the charged lepton that it 
produces in the detector~\cite{Battistoni}.
The enhancement is significant only for lower energies and for 
$|\cos(zenith)| \approx 0.2$, within 12 degrees from the horizontal 
direction. Since the angle between the neutrino
and the charged lepton it produces in the detector is expected 
to be $\sim$~60 degree below 0.5~GeV, the horizontal enhancement 
cannot be resolved.

\subsection{Pathlength distributions}

\begin{figure}[!tbh]%22
\centerline{\psfig{figure=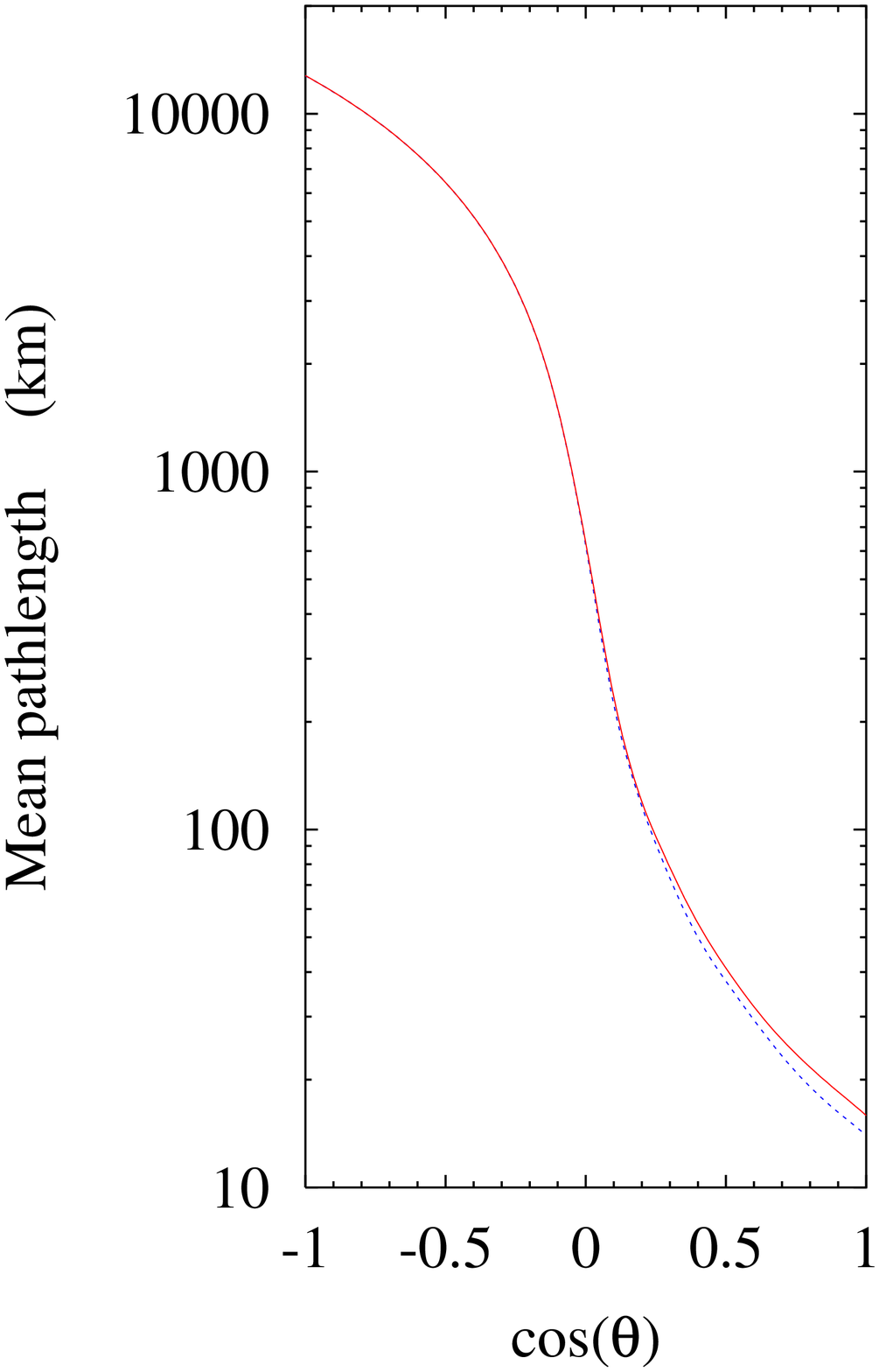,width=4.3cm} \hspace{.5cm}
{\psfig{figure=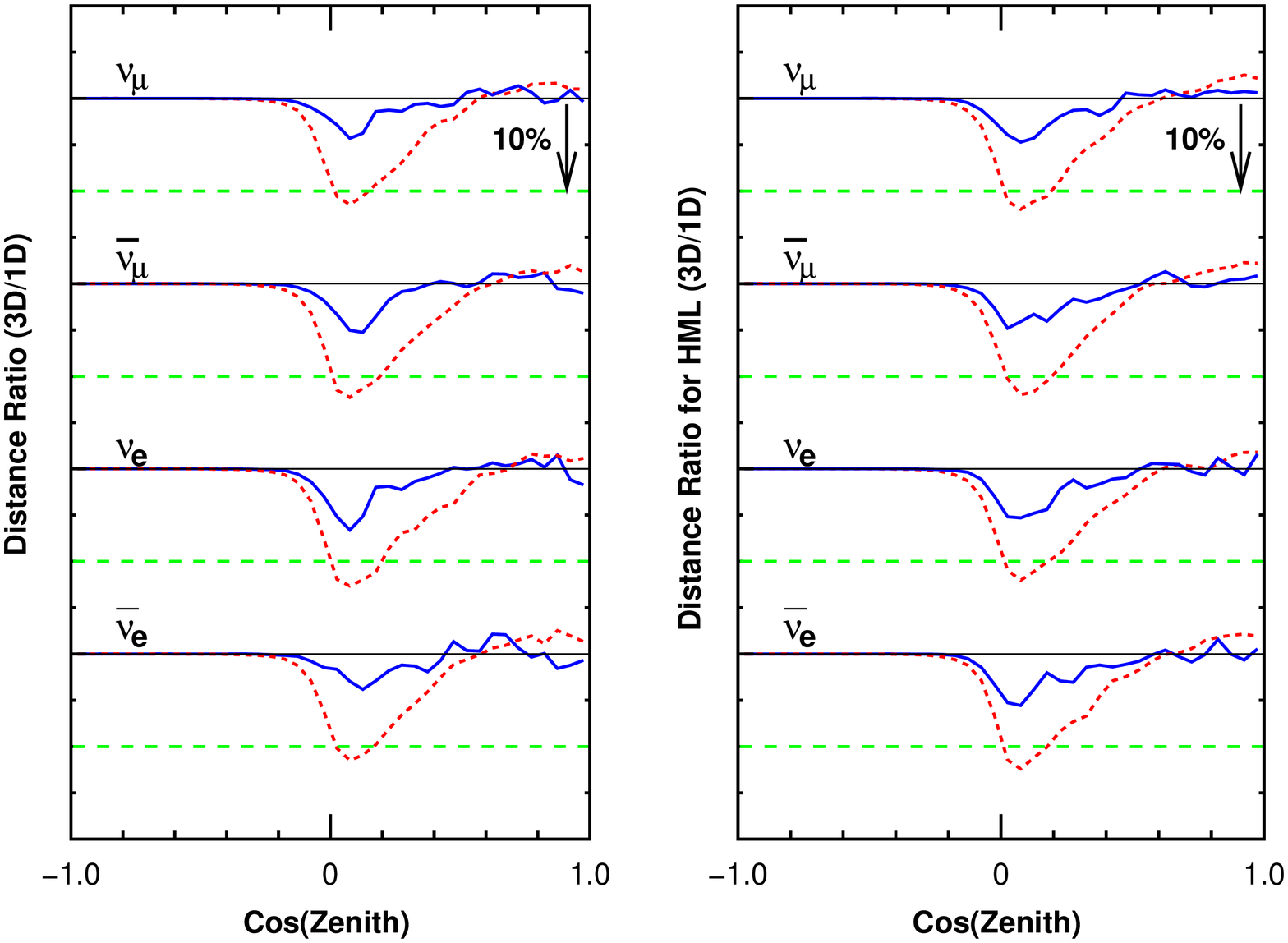,width=8cm}}}
\caption{a) Mean pathlength of neutrinos with $0.3<E_\nu<2$~GeV
in a 1D calculation~\protect\cite{pathlength}.
b) Ratio of the median path length in 3D calculation
to that in 1D calculation both for Super-K and North America.
Solid lines show the ratio for 1~GeV neutrinos, and dotted lines for
0.3~GeV neutrinos.
$\cos\theta = 1$ denotes the downward direction for neutrinos.
}
\label{distance-ratio}
\end{figure}

A key element of the interpretation of
the evidence for oscillations is the pathlength-dependence
of the atmospheric neutrino beam.  To a first approximation,
the distribution is a delta function equal to the distance
from the detector along the direction of the neutrino to a shell at
approximately 15 kilometers altitude.  The next step was to calculate
the pathlength distribution corresponding to the distribution
of production depths still assuming the neutrino is along the
direction of the primary cosmic ray that produced it~\cite{pathlength}.
This is the approach used so far in the analysis of the Super-K
data, and the mean value of the pathlength in
the 1D calculation~\cite{pathlength} is shown 
in Fig.~\ref{distance-ratio}a.  
What is still lacking is a treatment of the pathlength
distribution with the full three-dimensional geometry.  In particular,
the excess of low energy neutrinos near the horizon in the 3D
calculation arises in part from neutrinos produced closer to the detector
than neutrinos from the same direction in the 1D calculation.   
This is possible because the more nearly vertical nearby primaries
can contribute neutrinos at zenith angles that would require
larger pathlengths if the primary and the neutrino direction were
the same.  The nearby case just described is
not fully compensated by larger angle primaries producing smaller angle
neutrinos.  This is a consequence of the $\cos\theta_{cr}$ projection
factor in Eq.~\ref{3dyield}.

We compare the path length calculated in the 1D and 3D calculations,
converting the median production height to the path length 
by a simple relation:
\begin{equation}
\ell = \sqrt{(h^2+2R_\oplus h) + (R_\oplus \cos\theta)^2} - R_\oplus \cos\theta~~.
\end{equation}
We show the ratio of the two path lengths (3D$/$1D) as a function
of $\cos(\theta)$ in Fig.~\ref{distance-ratio} from Ref.~\cite{Honda3D}.
In this comparison an average over azimuth has been made.
Near the horizon the production distance of 0.3~GeV 
neutrinos is $\sim$~10~\% smaller
for 3D calculation than for 1D.
For 1~GeV the difference decreases to $\lesssim$ 5~\%.
%as is expected from the comparison of production height. 
%Honda-san---I do not understand this phrase???
As there is not a visible difference between the plots for
Kamioka and for high-latitude
North America, we may conclude that there is almost no geomagnetic 
latitude dependence of the path length differences.

\section{Atmospheric muons}

In this section we review the use of muon flux as a constraint 
on the atmospheric neutrino calculations.  We focus on the low
energy region, where pions give the dominant contribution
to neutrinos as well as to muons.

\begin{figure}[!htb]%23
\centerline{\psfig{figure=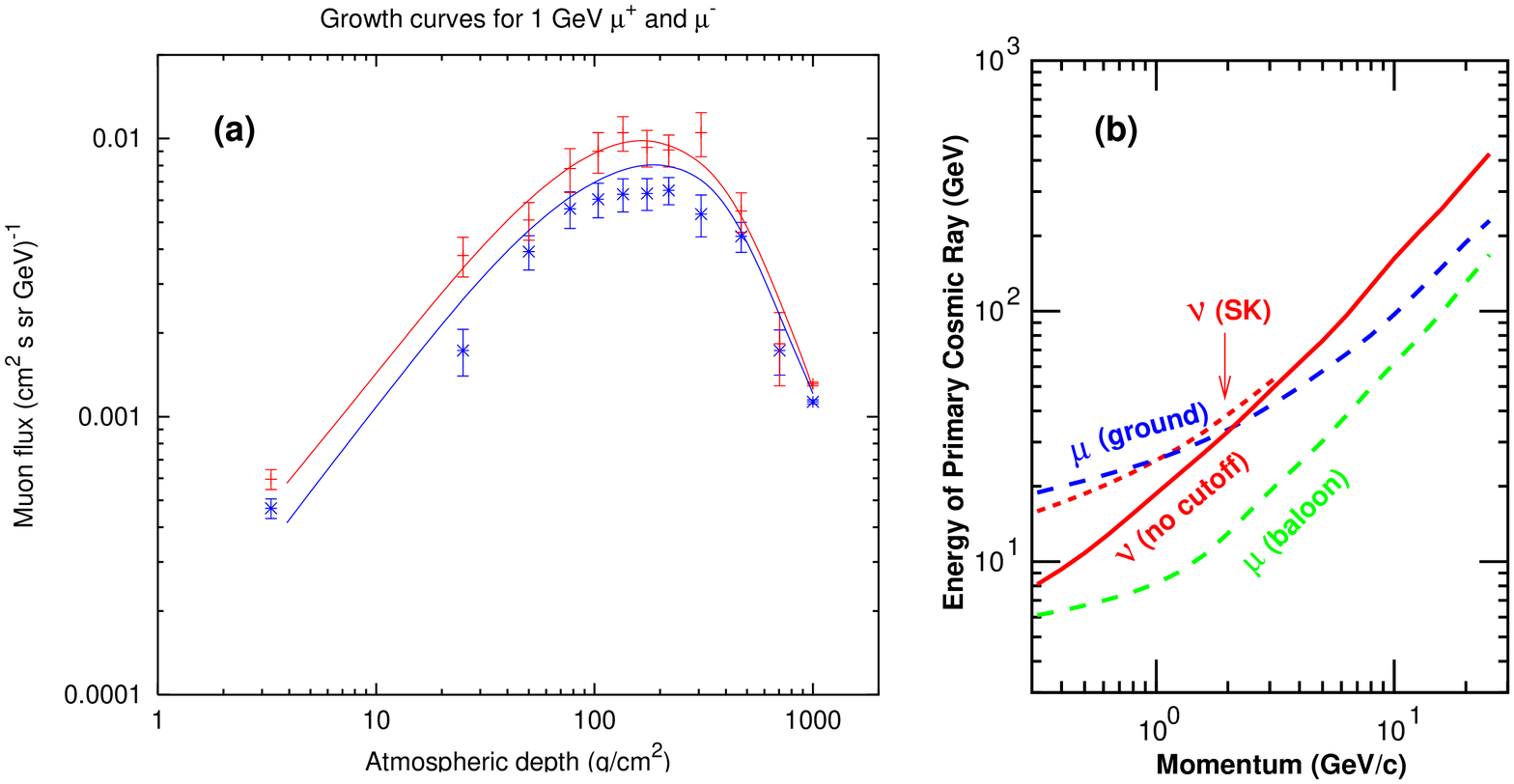,width=12cm}}
\caption{a. Growth curve of muons. 
The data are those of Ref.~\protect\cite{CAPmu}
and lines are 3D calculations from Ref.~\cite{Engmu}.
b. Median primary energy per nucleon for muons and 
neutrinos (see text).
}
\label{Fig-mu-growth}
\end{figure}

Fig.~\ref{Fig-mu-growth}a shows the growth curve of the 
vertical atmospheric 
muon flux for $E_\mu\approx 1$~GeV as a function of
atmospheric depth $X\,({\rm gm/cm}^2)$. The flux increases with depth 
from top of atmosphere down to $X\sim 100\;{\rm g/cm^2}$.
For $X \ll \lambda_N \approx 80$~gm/cm$^2$ (where $\lambda_N$
is the nucleon interaction length) the increase is proportional to $X$.
In the region of 100--200~g/cm$^2$, the attenuation of the primary
cosmic rays and the increase of the column density compensate each other,
and the muon flux is almost constant.
Below 200~g/cm$^2$, the muon flux decreases rapidly as muon
production decreases and muon decay and energy-loss become important.
Since decay probability and the importance of energy loss depend
on energy, the growth curve depend somewhat on energy, but
the qualitative behavior is unchanged.

Fig.~\ref{Fig-mu-growth}b shows the median primary energy per
nucleon for vertical muons (dashed lines) and neutrinos
averaged over all angles (heavy solid
and dotted curves).  
Comparison of response curves for muons at the ground
and at float altitude (typically $<5$~g/cm$^2$) shows the effect
of muon energy loss in the atmosphere.  Muons typically lose
$\approx 1.8$~GeV propagating through the atmosphere, and
most muons produced with this energy decay before reaching the
ground.  On the other hand, high in the atmosphere the muons
have not lost energy and are closer in energy to the primaries.
For neutrinos, the solid line is for Kamioka site, and the dotted line 
is the result that would be obtained in the absence of a
geomagnetic cutoff.
The response curve for north America or Gransasso would be 
between the solid and dashed lines.  Both muon charges
($\mu^+, \mu^-$), and all neutrinos species ($\nu_\mu$, $\bar\nu_\mu$, 
$\nu_e$, $\bar\nu_e$) were summed for these curves.

Measurements of muons in the cosmic radiation date back to the early days
of the subject.  A wealth of information on this and other
relevant topics is to be found in the comprehensive work of
Grieder~\cite{Grieder}.  There is also a recent study of
muons with $E\gtrsim10$~GeV at ground level in which the
authors have corrected the data for different observing
conditions~\cite{Hebbeker}.  A thorough summary is also
given in Ref.~\cite{Naumov}.  In the past decade, several
groups have measured the flux of muons on the ground and during
ascent to float altitude~\cite{MASS,CAPmu,HEAT,bess-sanuki}, and
several groups~\cite{Hondanew,Fior,Liu,Engmu,Battistonimu} 
have compared their calculations to some or
all of this data.
  
Several issues arise in comparing calculations with muon data.
One is whether to use the primary spectrum measured during 
the muon flight (if available) or to use some kind of best
fit or preferred spectrum.  For example, 
Battistoni et al.~\cite{Battistonimu} get generally
good agreement with the Caprice muon data~\cite{CAPmu}
using the primary spectrum
of Ref.~\cite{AGLS}, which is higher than that measured the
Caprice detector~\cite{CAPRICE} at the same time as the
muon measurement was made.  They note that
their normalization would have been generally lower than
the muon data had they used the Caprice primary spectrum.
On the other hand the CAPRICE primary
spectrum is used in Ref.~\cite{Engmu} to obtain
the results shown in Fig.~ref{Fig-mu-growth}a.  That calculation
uses the interaction model Target~2.1, which gives higher
neutrino fluxes at Soudan than the calculation with 
FLUKA~\cite{Battistoni}.  Since the muon fluxes are also measured
with no geomagnetic cutoff, the origin of the difference
between these two results is presumably the same--a difference
in pion production in interactions of $E_N<10$~GeV.

\begin{figure}[!htb]%24
\centerline{
\psfig{figure=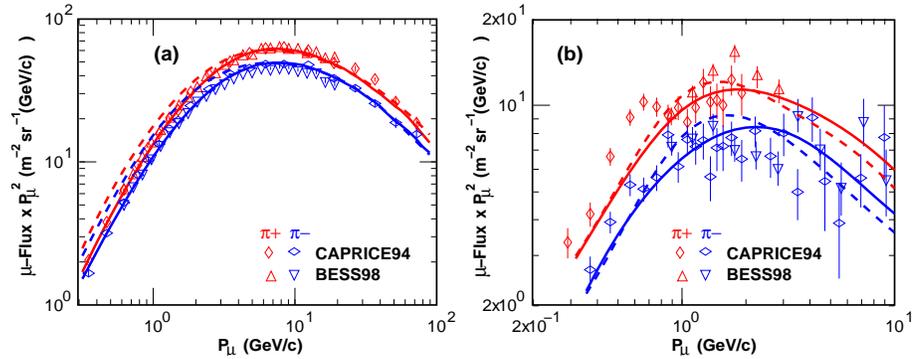,width=12cm}
}
\caption{
The flux of positive and negative muons observed at Lynnlake
and corresponding calculations.  Left panel refers to measurements
at the ground~\protect\cite{CAPground,bess-motoki}; right panel to
float altitude~\protect\cite{CAPmu,bess-sanuki}.
Thick solid and dotted lines show the positive and negative pion spectra 
calculated based on Dpmjet3 interaction model~\protect\cite{Hondanew}.
Thin solid and dotted lines show the positive and negative pion spectra 
calculated based on Fritiof 1.6 interaction model as used 
in~\protect\cite{Honda}.
}
\label{Fig-lynn-bess-caprice}
\end{figure}

Fig.~\ref{Fig-lynn-bess-caprice}~\cite{Hondanew} compares calculations
with data of two measurements both made in summer at Lynn Lake in Northern
Canada where the geomagnetic cutoff is negligible.
The Caprice muon data~\cite{CAPmu,CAPground} was obtained in 1994, while the
BESS muon data is a compilation of observations made in 1997, 1998, 
and 1999~\cite{bess-sanuki,bess-motoki}.  The neutron  
monitor count rates for both experiments (averaged over the
three dates in the case of BESS) are similar to that of 1998 
when BESS and AMS measured the primary cosmic ray spectrum
(see Fig.~\ref{modulation}).  The left panel shows a comparison
with the muon fluxes at the ground.  The comparison at float
altitude is shown on the right.  It is noteworthy that the
two measurements of the muon flux at the ground are in
better agreement with each other than the corresponding two
measurements of the primary proton spectrum~\cite{CAPRICE,BESS}.
Despite relatively long exposures at float altitude, the
statistics are poor because of the low flux.

Finally, we note that it may also be useful to compare high-statistics 
ground level measurements made a different geomagnetic locations.
Ref.~\cite{CAPground} compares
a measurement at Lynn Lake ($R_c\approx 0.5$~GV) with one at Ft. Sumner
($R_c\approx 4.2$~GV).  Ref.~\cite{bess-motoki} compares their
ground level measurement at Lynn Lake with a measurement at
Tsukuba, Japan, which has a higher cutoff of $11.4$~GV.  
The charge ratios measured at the
locations with a significant geomagnetic cutoff
fall below those measured at Lynn Lake for $E_\mu<1$~GeV.
This is expected qualitatively for several reasons.
Helium is relatively more important at high cutoff
because the energy per nucleon at cutoff is approximately
half that of protons, and helium produces a charge ratio of unity.
In addition, when the cutoff is high the low energy muons must
come more from the central region of interactions
above cutoff, and the charge ratio is lower in the
central region.  Finally, there is also some
contribution from the east-west effect since vertical
muons of opposite sign are bent in opposite directions~\cite{Utah}.

We summarize this section as follows.
Because of decay and energy loss, low energy 
muons at the ground carry less information about neutrino production
than those at high altitude where most of the neutrinos
are produced.  In addition, the variation of atmospheric
structure from barometric and seasonal
changes is significant and must be accounted for.
At float altitude, the muons reflect rather
directly the properties of pion production
and primary cosmic-ray intensity since only one
interaction is involved, but this is above the depth where
most muons and neutrinos are produced and the flux
is very low.  On the other hand,
the peak of the grown curve has so far been studied only
during ascent where the balloon rises rapidly
through the atmosphere, so that statistical and systematic
uncertainties are larger than on the ground or at float.
In addition, at all altitudes, the fraction of $\sim$GeV muons
passing through a detector
is always much smaller than the number of muons that were
produced above it~\cite{Engmu}.  Thus calculating the 
muon flux is more delicate than calculating the neutrino
flux which is an integral over all depths.

\section{Uncertainties in the Calculated Neutrino Fluxes}

Remaining uncertainties in the flux of
atmospheric neutrinos come mainly from the primary spectrum
and from the treatment of hadronic interactions. 
In addition, there remain uncertainties of a
technical nature related to remaining approximations
in present calculations.  The uncertainty in the signal
for a given experiment depends on its response function,
which determines the distribution of primary energy responsible
for the signal.  Key examples are shown in Fig.~\ref{response1}.

\subsection{Uncertainties from primary spectrum}
An estimate~\cite{spectrum} 
of the uncertainty in the primary spectrum in the
light of recent measurements by BESS~\cite{BESS} 
and AMS~\cite{AMS} is
$\pm5$\% below $100$~GeV/nucleon increasing to $\pm10$\% at $10$~TeV/nucleon.  
This is
based on a fit (Eq.~\ref{Todor} 
to BESS and AMS data alone and assumes a power-law to extrapolate
beyond the upper range of those measurements ($\approx 100$~GeV).

A more proper estimate of the uncertainty in the primary
spectrum uses all valid measurements,
including those with larger quoted uncertainties, to estimate the systematic
uncertainty in the primary spectrum.  
Fig.~\ref{Fig-primary} shows a collection of 
measurements for protons and helium.
Although there are now several recently reported measurements
below 100 GeV, there is a striking lack of recent data in 
the TeV range, important for $\nu$-induced upward muons.
Even assuming the high-energy data should be renormalized down to
connect with the low-energy data, the measurements cover a range of
$\pm20$\% below 100~GeV and $\pm30$\% above.

\subsection{Uncertainties from hadronic interactions}
It is more difficult to quantify the uncertainties due to
hadronic interactions.  As shown in Fig.~\ref{Raeven}, large
regions of phase space are not measured, and models are needed
to interpolate and extrapolate into the unmeasured regions.
Study of Fig.~\ref{hondaplot} show a scatter of 20-25\% even
if we consider the data of Ref.~\cite{Lundy} as an outlier.
On the other hand, this treats the different inclusive cross
sections as independent of each other, when in fact they
are interrelated by energy conservation.  

Combining the
systematic uncertainties in primary spectrum and in pion production
as if they were independent statistical uncertainties, we would
therefore estimate an overall theoretical uncertainty of $\pm$20 to 25\%
in the energy range relevant for contained neutrino interactions
and somewhat larger at higher energy.

\subsection{Other sources of uncertainty}

Calculations usually assume 
a spherical earth with the surface at sea level.
When parent mesons enter the ground before decaying they
loose energy rapidly by ionization or the nuclear interactions
and produce only very low energy neutrinos.
Therefore, when there is a high mountain over the neutrino detector,
the neutrino flux is reduced to some extent.  The effect is
negligible except near the vertical and then only for decay products
of muons.  The reduction has been estimated~\cite{Honda2000} 
as a 2--3~\% effect
for $\nu_e$ in the 10-100 GeV range
due to the Ikenoyama mountain over Super-Kamiokande,
which has a peak at approximately 1300~m above sea level.

Calculations are normally made for an average atmosphere.
While pressure and seasonal variations have a noticeable effect
on surviving muon fluxes at the ground, they are small for 
for neutrinos.  Variations at the per cent level have been
estimated in Ref.~\cite{Honda2000}.

\subsection{Comparison to measurements}

A consistent picture emerges from comparison of a variety of
data to calculations of the flux of atmospheric neutrinos.   
All analyses emphasize comparison to ratios, which include
$\nu_e/\nu_\mu$ and up/down for contained events and
stopping/throughgoing and vertical/horizontal for neutrino-induced
upward muon.  As a consequence, an offset in normalization does
not affect the conclusions about oscillations, which are robust.

\begin{figure}[!thb]%25
\centerline{{\psfig{figure=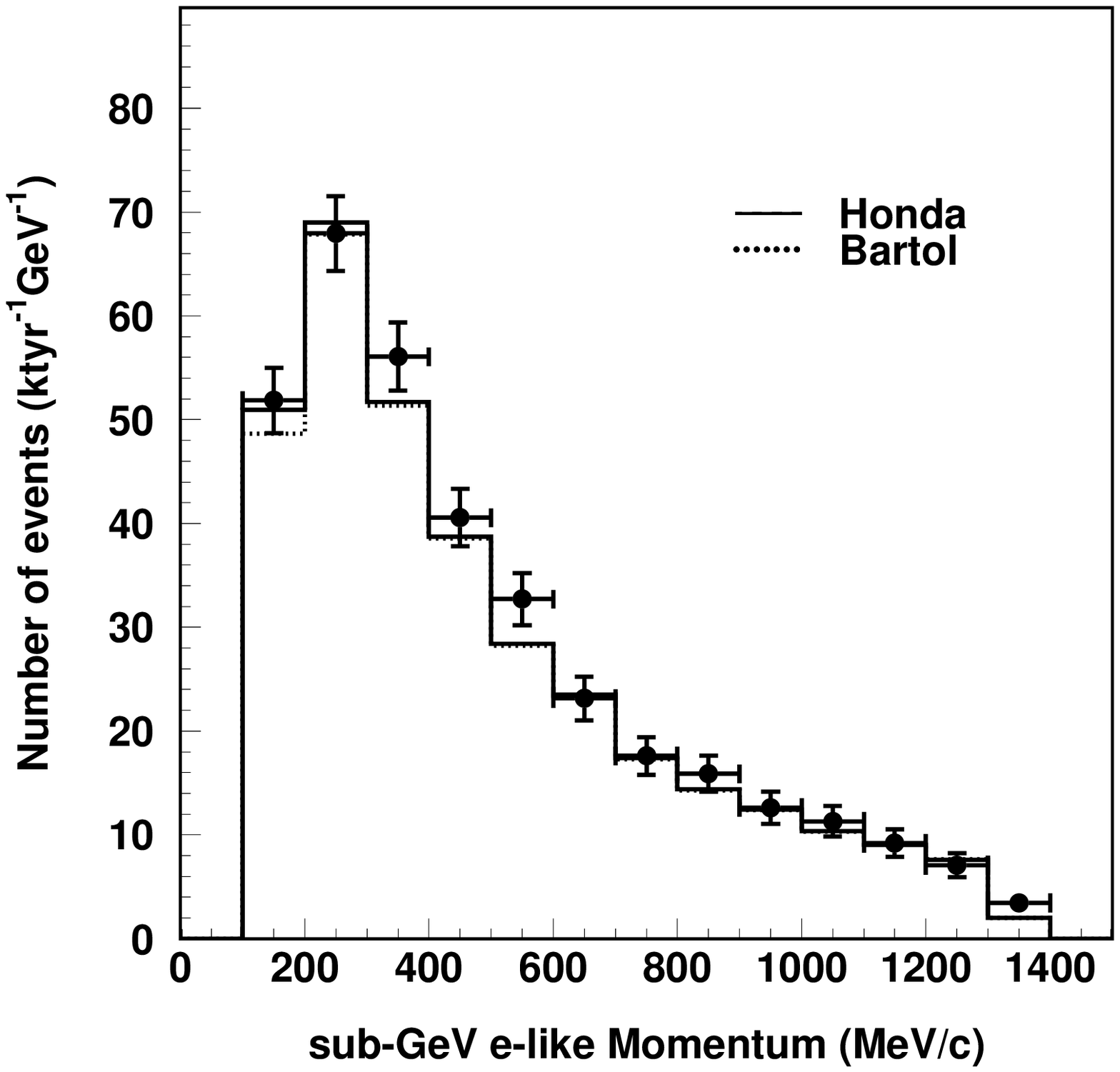,width=6.cm} \hspace{5mm}
{\psfig{figure=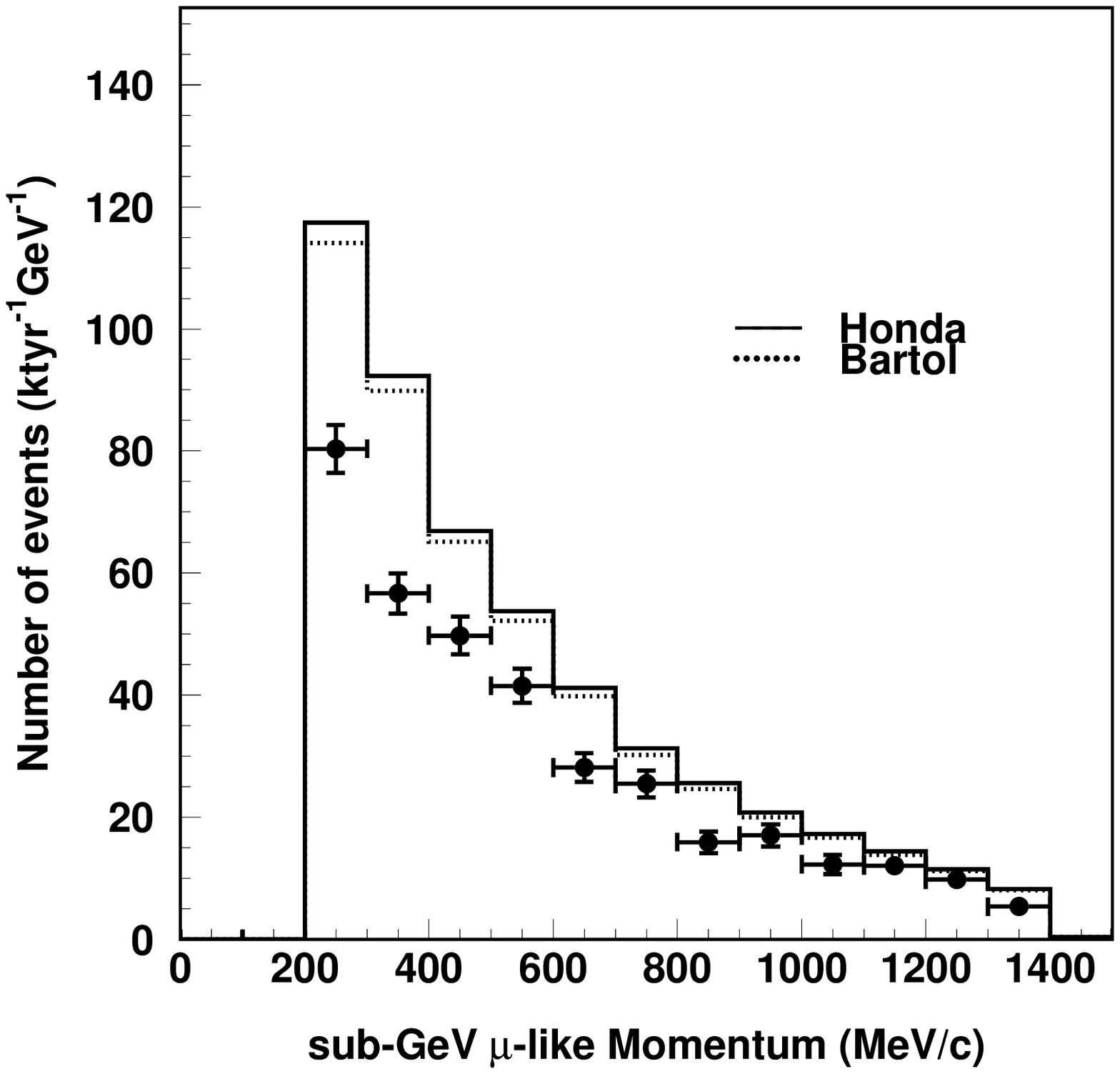,width=6.cm}}}}
\caption{Sub-GeV neutrino spectrum.
}
\label{subGeV}
\end{figure}

Nevertheless, it is also interesting and informative to compare the
measured neutrino flux as directly as possible in an absolute
fashion with the calculations.
Fig.~\ref{subGeV} compares the neutrino fluxes
measured at Super-K with the two calculations~\cite{Honda,AGLS} 
that have been used
extensively as input to the detector simulation.
The comparisons are made as a function of visible energy by
the Super-K group~\cite{Kajita-private}.
The magnitudes are consistent with what is expected
for oscillations in this energy range that involve
$\nu_\mu$ and $\nu_\tau$ but not $\nu_e$; namely, there is
a deficit of $\nu_\mu$, whereas the measured flux of electron
neutrinos is consistent with the calculation.

Systematic comparisons of more recent calculations with data
are not yet available.
Based on comparisons between old and new 
calculations such as those shown in Fig.~\ref{tar-tom}, however, 
we expect
that the new calculations will give significantly lower predicted
neutrino fluxes.  In Refs.~\cite{Hamburg,FLUKA} this will be a 
consequence of lower pion production, whereas in the case
of Ref.~\cite{Hondanew} the primary spectrum is lower than
before~\cite{Honda}.  We expect the
predictions may be reduced by as much
as 15 to 20\%, without changing the ratios significantly.
Possible resolutions include uncertainties
in the neutrino cross sections~\cite{Sartogo,Nuint01} or
systematic uncertainty in primary spectrum (despite the 
agreement between BESS~\cite{BESS} and AMS~\cite{AMS}
or higher pion production (despite new calculations)
or some combination of the above.

%%%%%%%%%%%%%%%%%%%%%%%%%%%%%%%%%%%%%%%%
\begin{figure}[!thb]%26
\centerline{\psfig{figure=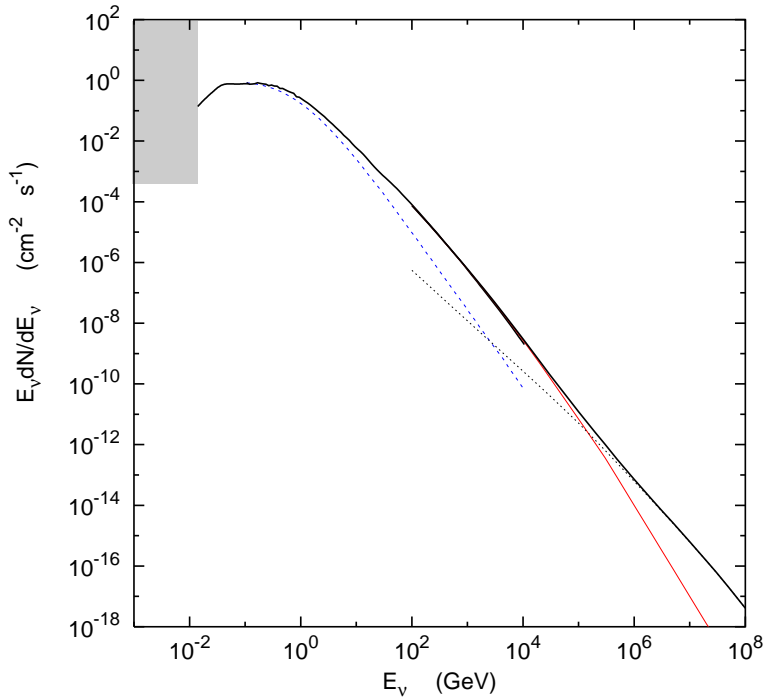,width=11cm}}
\caption{\label{global}
Global view of the neutrino spectrum: vertical flux of $\nu_\mu+\bar{\nu}$
(heavy solid line); $\nu_e+\bar{\nu}_e$ (dashed line); prompt neutrinos
(dotted line);  $\nu_\mu+\bar{\nu}$ from pions and kaons (thin solid line
at high energy).  
}
\end{figure}
%%%%%%%%%%%%%%%%%%%%%%%%%%%%%%%%%%%%%%%%

\section{Global view of the neutrino spectrum}

Figure~\ref{global} is a global view of the omnidirectional
neutrino flux from the MeV range up
to 100 PeV.  Below 15-18~MeV solar neutrinos dominate,
as indicated by the shaded region.  
It has been pointed out~\cite{SNe1,SNe2,SNe} that
there is potentially a window between $\sim15$ and $50$~MeV
where the accumulated diffuse flux from supernovae in the
Universe could be comparable to or larger than the flux of
atmospheric neutrinos.

The heavy solid line
shows the flux of atmospheric $\nu_\mu+\bar{\nu}_\mu$ integrated over all
directions.  The dotted lines show the flux of $\nu_e+\bar{\nu}_e$.
Where geomagnetic effects are important (below several GeV), 
we show the integrated flux at Kamioka.
With the discovery of oscillations, the flux of muon neutrinos
is suppressed in the energy region below 100 GeV.  We used
the Super-K oscillation parameters~\cite{evidence} assuming
$\nu_\mu\leftrightarrow\nu_\tau$ to relate the production spectrum
of $\nu_\mu+\bar{\nu}_\mu$ as calculated in Ref.~\cite{AGLS}
to the flux at the detector.  Thus for
$E_\nu < 10$~GeV the $\nu_\mu/\nu_e$ ratio decreases toward 1.

For $E_\nu\sim100$~GeV and beyond, most neutrinos come from decay
of kaons until the energy is so high that prompt neutrinos dominate.  
The dotted line at higher energy is the flux of prompt
neutrinos ($\nu_e+\bar{\nu}_e\approx\nu_\mu+\bar{\nu}_\mu$)
from decay of charm, as estimated in the RQP model of
Ref.~\cite{Bugaev}.  In this model, the charm crossover is
at $\sim 3$~TeV for electron neutrinos and around $100$~TeV
for muon neutrinos.  The thinner (lower) solid line above $100$~TeV
is the non-prompt contribution to the flux of $\nu_\mu+\bar{\nu}_\mu$.
In the TeV region and above the atmospheric neutrinos become
a foreground for high energy astrophysical neutrinos, and the
uncertainty in the level of prompt neutrinos 
becomes an important consideration in this context.  
 
\section{Acknowledgments}  We are grateful to 
Giles Barr, Giuseppe Battistoni,
Ralph Engel, Frank Jones, T. Kajita, K. Kasahara, Paolo Lipari,
S. Midorikawa, Teresa Montaruli, T. Sanuki and Todor Stanev for their
assistance and advice as we prepared this review.  
The data plotted in Fig.~\ref{modulation}
are from research supported by
National Science Foundation Grant ATM-9912341 at the University
of Chicago.  One of us (TKG)
gratefully acknowledges the hospitality of the Institute for Cosmic
Ray Research where most of this review was prepared.  The work of
TKG is supported in part by the U.S. Department of Energy under
DE-FG02 91ER 40626.  

\section{LITERATURE CITED}

\end{document}